\documentclass[a4paper,10pt]{aa}

\usepackage{graphicx}
\usepackage{txfonts}
\usepackage{natbib}
\usepackage[normalem]{ulem}

\usepackage{subfigure}

\title{An incisive look at the symbiotic star \object{SS Leporis}\thanks{Based on observations made with the VLTI European Southern Observatory telescopes obtained from the ESO/ST-ECF Science Archive Facility.}}
\subtitle{Milli-arcsecond imaging with PIONIER/VLTI}

\author{N. Blind \inst{1} \and H.M.J. Boffin \inst{2} \and J.-P. Berger \inst{2}  \and J.-B. Le Bouquin \inst{1} \and A. M\'erand \inst{2} \and B. Lazareff \inst{1} \and G. Zins \inst{1}}

\institute{UJF-Grenoble 1/CNRS-INSU, Institut de Plan\'etologie et d'Astrophysique de Grenoble (IPAG) UMR 5274, Grenoble, France \\
e-mail: \texttt{nicolas.blind@obs.ujf-grenoble.fr}
\and
European Southern Observatory, Casilla 19001, Santiago 19, Chile   \\ e-mail: \texttt{hboffin@eso.org}}

\date{Received 7 September 2011 / Accepted 20 November 2011}

\begin{document}

\abstract
	{Determining the mass transfer in a close binary system is of prime importance for understanding its evolution. SS Leporis, a symbiotic star showing the Algol paradox and presenting clear evidence of ongoing mass transfer, in which the donor has been thought to fill its Roche lobe, is a target particularly suited to this kind of study.}
	{Since previous spectroscopic and interferometric observations have not been able to fully constrain the system morphology and characteristics, we go one step further to determine its orbital parameters, for which we need new interferometric observations directly probing the inner parts of the system with a much higher number of spatial frequencies.}
	{We use data obtained at eight different epochs with the VLTI instruments AMBER and PIONIER in the H- and K-bands. We performed aperture synthesis imaging to obtain the first model-independent view of this system. We then modelled it as a binary (whose giant is spatially resolved) that is surrounded by a circumbinary disc.}
	{Combining these interferometric measurements with previous radial velocities, we fully constrain the orbit of the system. { We then determine the mass of each star and significantly revise the mass ratio. The M giant also appears to be almost twice smaller than previously thought. Additionally, the low spectral resolution of the data allows the flux of both stars and of the dusty disc to be determined along the H and K bands, and thereby extracting their temperatures.}}
	{We find that the M giant actually does not  \protect{\it stricto sensus} fill its Roche lobe. The mass transfer is more likely to occur through the accretion of an important part of the giant wind. We finally rise the possibility for {  an enhanced mass loss from the giant, and we show that an accretion disc should have formed around the A star}.}
	

\maketitle

\section{Introduction} \label{part:intro}

Symbiotic stars are interacting binaries composed of a hot star accreting material from a more evolved red giant companion. They are excellent laboratories for studying a wide spectrum of poorly understood physical processes, like the late stage of stellar evolution, the mass loss of red giants, and the mass transfer and accretion in binary systems \citep{mikolajewska_2007}. Their study has important implications for a wide range of objects, like Type Ia supernovae, barium stars, the shaping of planetary nebulae, and compact binaries like cataclysmic variables \citep{podsiakolski_2007}. 

SS Leporis (17 Lep; HD 41511; HR 2148) is a prime example of such a long-period interacting system, even though it does not belong to the most common symbiotic systems, because the hot star is not the usual compact white dwarf. As such, SS Lep is a symbiotic system in the first phase of mass transfer, while most symbiotic stars are in their second episode of mass transfer, following the first one that produced the white dwarf. 

SS Lep has been known for many decades to present symbiotic features, and its optical spectrum shows at least three components \citep{struve_1939, molaro_1983, welty_1995}. The spectral lines of an A star are largely obliterated by shell features that dominate at shorter wavelengths, while an M star spectrum becomes increasingly obvious at longer wavelengths. \citet{welty_1995} estimated an M4 III spectral type for the cool companion, while even earlier types have been estimated by previous authors. The shell is absorbing light primarily from the A star, indicating some mass loss from the hotter star. The system, however, presents the so-called Algol paradox, as the most evolved star is also the least massive, which implies that the cool star must have lost a large quantity of matter and that the hot companion has accreted part or most of it. Moreover, the regular ``outbursts'' \citep{struve_1930, welty_1995} and the UV activity \citep{polidan_1993} of the A star shell are clear testimony t ongoing mass-transfer episodes. From interferometric observations, \citet{verhoelst_2007} inferred that the mass transfer occurs because the M giant fills its Roche lobe. 

The binary system is additionally surrounded by a large circumbinary dust disc and/or envelope \citep{jura_2001}. Interferometric observations confirmed this fact by revealing its presence in the inner part of the system \citep{verhoelst_2011}, further noticing that the structure must be in a disc-like geometry to be compatible with the low extinction towards the central star. \citet{jura_2001} suggest that the circumbinary disc contains large grains that are formed by coagulation and, based on the large and rather unique 12 $\mu$m excess of SS Lep despite its rather low luminosity,  that the disc may be losing mass by a wind at a rate of $8\times10^{-9}$ M$_\odot$yr$^{-1}$.

The orbital characteristics and circumbinary disc of SS Lep very closely resemble those of the post-AGB binaries with stable, Keplerian circumbinary dust discs \citep{vanwinckel_2003}, and, as such, SS Lep may be considered as a system linking binary M giants and post-AGB systems -- the M-giant should indeed very soon evolve into a post-AGB star. In those post-AGB binaries, the spectra are rich in crystalline features while the spectrum of SS Lep appears entirely amorphous, which, if a link is indeed in order, would imply further dust processing in the disc. Whether the disc can survive long enough if it is indeed losing mass through a wind is, however, still an open question. 

\citet{cowley_1967} argued that the system consists of a B9 V primary and an M1 III secondary in a 260-d eccentric orbit ($e=0.132$). She developed a scenario in which the secondary fills its Roche lobe near periastron and mass transfer proceeds for a short time thereafter. From spectra covering 3.5 orbits, \citet{welty_1995} proved this scenario unlikely, as their revised orbit provided a similar orbital period but a significantly reduced eccentricity  $e=0.024$. They estimated a mass ratio of $1/q = 3.50 \pm 0.57$, where the error was very likely severely underestimated given the poor fit of the single Mg II line they used to measure the radial velocity of the A star. 

Recently, \citet{leeuwen_2007} has reevaluated the parallax of SS Lep from Hipparcos data, obtaining $\pi = 3.59\pm0.31$\,mas, that is, a distance of $279\pm24$\,pc, so smaller than the previously and generally used value of $330^{+90}_{-60}$\,pc. The most recent parameters of SS Lep as collected from the literature until the present work, is presented in Table~\ref{tab1}.

We report here interferometric observations that have allowed model-independent image synthesis (Section \ref{part:image}) and a more precise modelling of the system (Section \ref{part:model}). Section \ref{part:binary} focuses on the binary by determining mainly the orbital parameters and the M star diameter. We finally discuss the mass transfer process in Section \ref{part:masstransfer}.

%
\begin{table}[t!]\label{tab1}
\caption{\label{tab:param} Previously estimated parameters of SS Lep}
\centering
 \begin{tabular}{lc|lcc}
\hline
\hline
                        & System &          & A star & M star   \\
\hline
  $d$ [pc]        & $279\pm24$ $^a$ &  {\it SpT}                       & A1V $^{b}$    & M6III $^{c}$ \\
  $P$ [d]          & $260.3\pm1.8$ $^b$     & $T_{\rm eff}$ [K]  & 9000 $^{c,d}$   & 3500 $^{c,d}$ \\
  $e$          & $0.024\pm0.005$ $^b$ &$\theta$ [mas]   & $0.53\pm0.02$  $^c$  & $3.11\pm0.32$  $^c$\\
  $i$           & $30^\circ \pm 10^\circ$ $^c$   &$R$ [${\rm R}_\odot$]        & $\sim$15 $^{c}$      & $110\pm30$ $^{c}$ \\
  $f(M)$& 	$0.261\pm0.005$ $^b$ &$F$ [\%] & $11\pm7$ $^c$     & $60\pm50$ $^c$ \\
  $1/q$            & $4\pm1$ $^{b,c}$  & $M$ [${\rm M}_\odot$] & 2$\sim$3 $^{a,b}$ & 0.4$\sim$1 $^{a,b}$ \\
\hline

\hline
 \end{tabular}
\tablefoot{$d$ is the distance, $P$ the orbital period, $e$ the eccentricity, $i$ the inclination, $f(M)$ the mass function, $q=M_M/M_A$ the mass ratio. For the stars, {\it SpT} is the spectral type, $T_{\rm eff}$ the temperature, $\theta$ the apparent diameter, $R$ the linear radius, $F$ the flux contribution at $2.2\,\mu m$, and $M$ the mass. References: (a) \citealt{leeuwen_2007}; (b) \citealt{welty_1995}; (c) \citealt{verhoelst_2007}; (d) \citealt{blondel_1993}.}
\end{table}
%

\section{Observations} \label{part:obs}

%
\begin{table}[t!]
\centering
\caption{VLTI Observation log of SS Lep  for the AMBER and PIONIER observations. \label{tab:obslog}}
\begin{tabular}{lrlll}
\hline \hline
Run & Date & Baselines & { Range} \\
\hline
A1 & 11-11-2008 & E0-G0-H0 & 15 - 130\,m \\
    & 13-11-2008 &  A0-G1-K0 & \\
A2 & 26-12-2008 &  A0-G1-K0 & 90 - 130\,m \\
A3 & 21-02-2009 & D0-G1-H0 & 15 - 75\,m\\
    & 28-02-2009 & E0-G0-H0 & \\
A4 & 07-04-2009 & D0-G1-H0 & 65 - 70\,m \\
\hline
P1 & 28-10-2010 & D0-G1-H0-I1 & 15 - 80\,m\\
    & 30-10-2010 & D0-E0-H0-I1  & \\
P2 & 29-11-2010 & E0-G0-H0-I1 & 15 - 70\,m  \\
P3& 07-12-2010 & D0-G1-H0-I1 & 45 - 80\,m \\
P4 & 22-12-2010 & A0-G1-K0-I1 &  45 - 130\,m\\
\hline
\end{tabular}
\end{table}
%
%
\begin{figure}[b!]
	\centering
	\includegraphics[width=.23\textwidth]{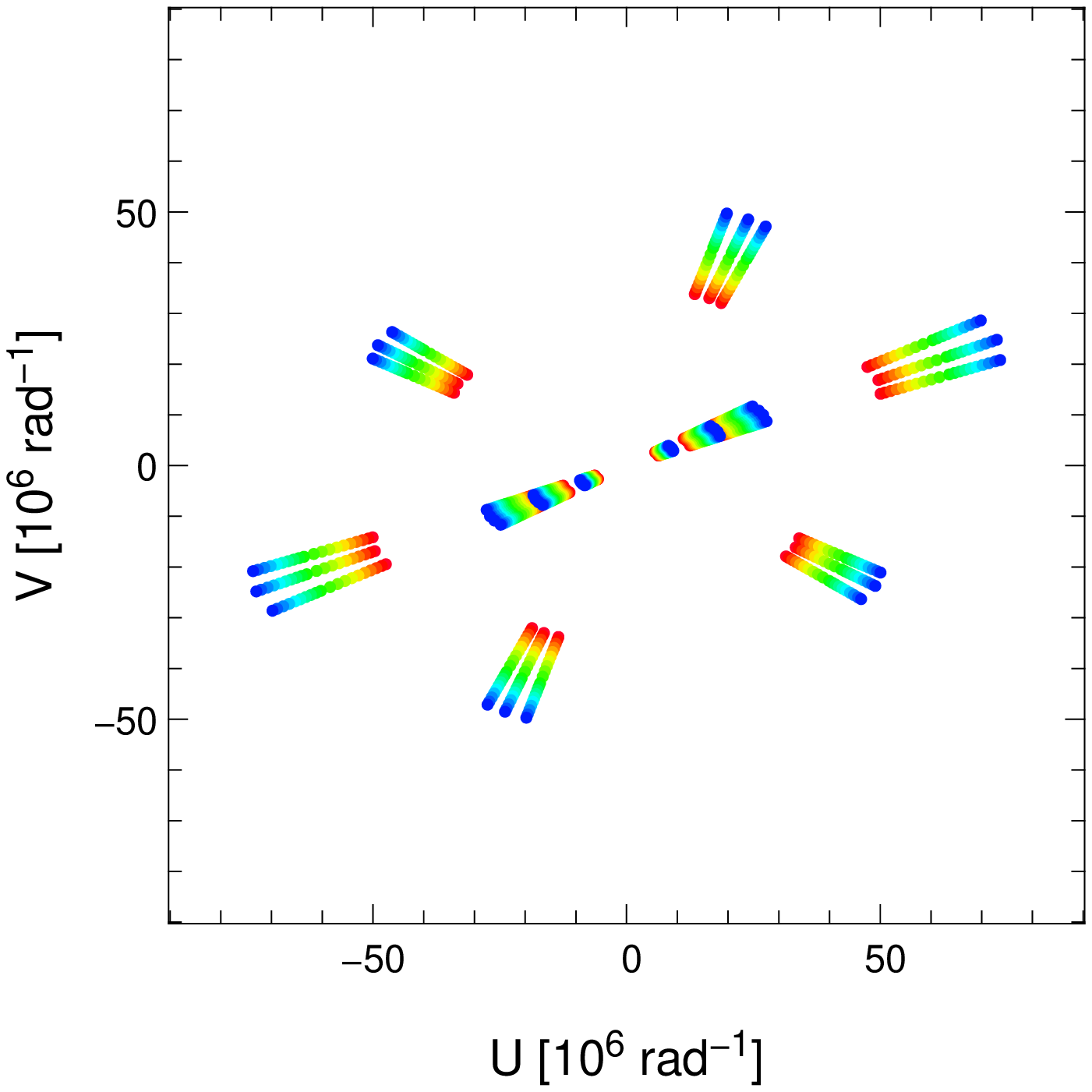}
	\includegraphics[width=.23\textwidth]{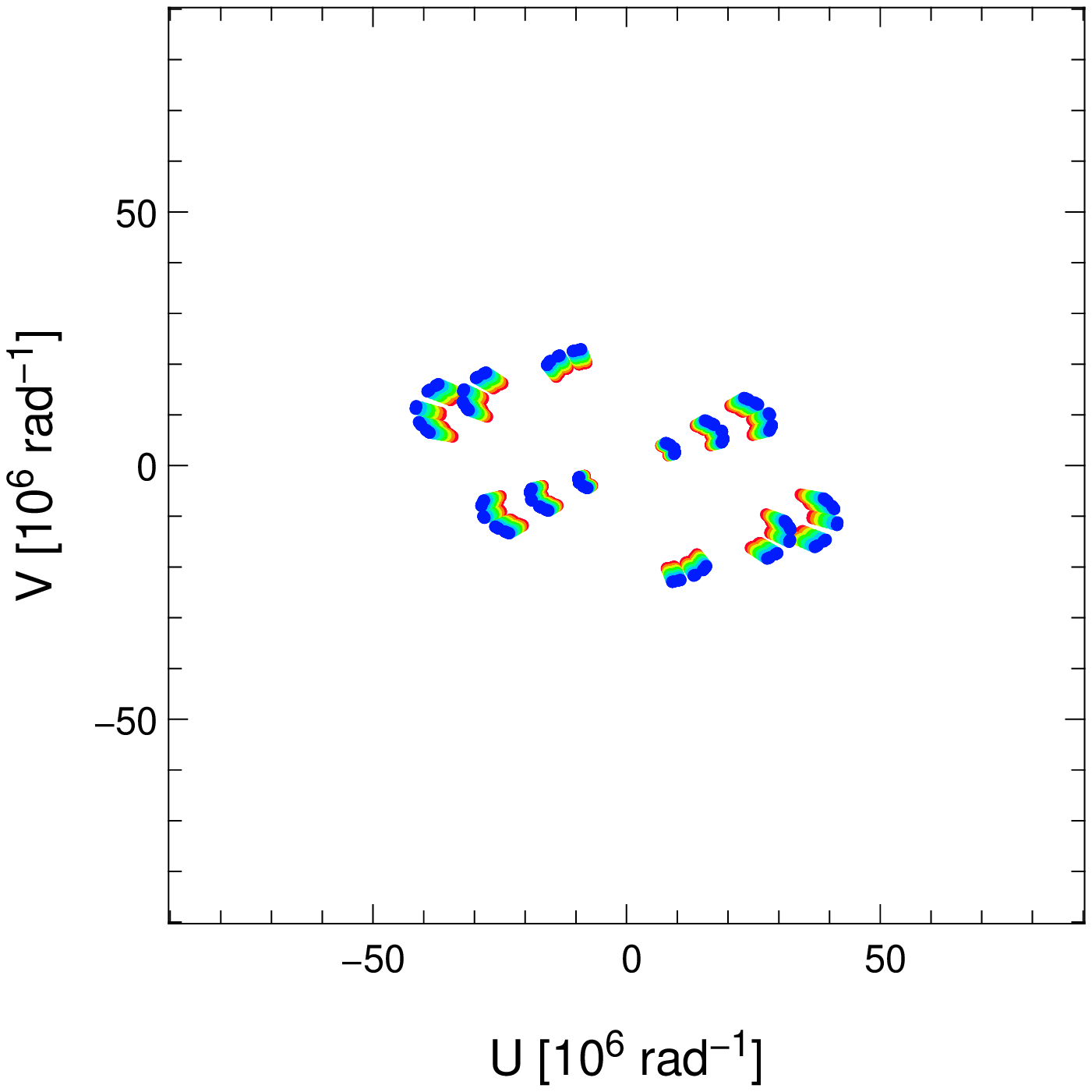}
	\caption{Typical {\it (u,v)-}plane coverage for AMBER (left, November 2008) and PIONIER (right, October 2010) observations. \label{fig:UV_plan}}
\end{figure}
%

Data were collected at the Very Large Telescope Interferometer \citep[VLTI;][]{haguenauer_2010} with the spectrograph AMBER \citep{petrov_1997} and the four-telescope visitor instrument PIONIER \citep{lebouquin_2011a}. All observations made use of the 1.8-m Auxiliary Telescopes. Table~\ref{tab:obslog} presents the observation log. Three data sets are the combination of two observations separated in time by a few days (seven to the maximum) to increase the {\it (u,v)-}plane coverage. We made sure that this brings valuable information without biasing the results, more specifically the estimation of the binary orientation.  Typical {\it (u,v)-}planes for AMBER and PIONIER observations can be seen in Fig. \ref{fig:UV_plan}.


\paragraph{AMBER data} We used archive AMBER data obtained during four different nights in a period of 200\,days (more than half an orbital period). They cover simultaneoulsy the J-, H-, and K-bands with a spectral resolution $R\sim35$. Even though J-band fringes have been properly recorded in several observations, we decided to discard them from the analysis of this paper since the data quality is significantly worse than for longer wavelengths. Raw visibility and closure phase values were computed using the latest public version of the \texttt{amdlib} package \citep[version 3;][]{malbet_2010} and the yorick interface provided by the Jean-Marie Mariotti Center.

\paragraph{PIONIER data} They were obtained in the H-band between the end of October 2010 and December 2010 during the commissioning runs of the instrument. We used the prism that provides a spectral resolution $R\sim40$, that is, six spectral channels across the H-band. Because these observations have been made during the commissioning, the {\it (u,v)}-plane coverage is still relatively poor for a whole night of observations but more complete than AMBER's. Data were reduced with the \texttt{pndrs} package presented by \citet{lebouquin_2011a}.\newline

The low spectral resolution multiplies the number of spatial frequencies, and brings a wealth of information (Sect. \ref{part:model}). This is especially true with AMBER for which the {\it (u,v)-}coverages were relatively poor.

%
\begin{figure*}
	\centering
	\includegraphics[angle=270, width=0.99\textwidth]{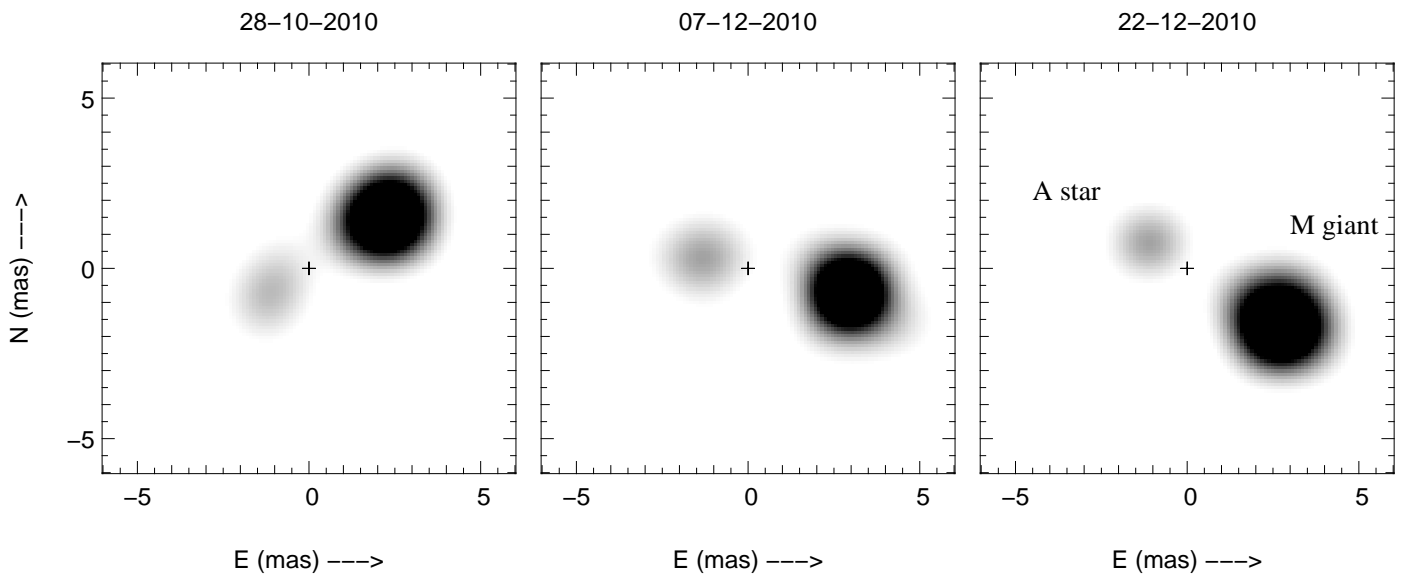}
	\caption{Model-independent image reconstruction of SS Lep obtained during the PIONIER runs P1, P2, and P4. The resolved M giant and the A star are clearly identified. The images are centered on the center of mass (central cross) as determined from Section \ref{part:masses}. The distortion of the giant in the image is most certainly due to an asymmetric PSF rather than to a definite tidal effect. Three faint artefacts are visible on the periphery of the image. \label{fig:SSLep_images} }
\end{figure*}
%

\section{Image synthesis} \label{part:image}

With its four telescopes, PIONIER provides six visibilities and four closure phases simultaneously, which allows a reliable image reconstruction for the four observations. We used the MIRA software from \citet{thiebaut_2008}. MIRA proceeds by direct minimisation of a penalised likelihood. This penalty is the sum of two terms: a likelihood term that enforces agreement of the model with the data (visibilities and closure phases), plus a regularisation term to account for priors. The priors are required to lever the possible degeneracies due to the sparseness of the spatial frequency sampling. We use here the ``total variation'' regularisation associated with positivity constraint as recommended by \citet{renard_2011}. The pixel scale is $0.25$ mas and the field-of-view is $200\times200$ pixels. The starting point is a Dirac function in $(0,0)$. We set the hyper-parameter to a low value of 100, so that the weight of the regularisation term is kept small with respect to the fit to the data. It brings some super-resolution, at the cost of an improved noise level in the image. We combined all the spectral channels to improve the $(u,v)$-plane coverage. That the image is indeed ``grey'' over the H-band is demonstrated in the next section.  The reconstructed images for runs P1, P2, and P4 are presented in Fig.~\ref{fig:SSLep_images}. Each image shows the binary nature of SS Lep, the separation being slightly smaller than 5\,mas. From one observation to the next, we can observe the rotation of the system.

The A star and its shell have an expected spatial extension of 0.5\,mas \citep{verhoelst_2007} so that we do not expect to resolve them with our VLTI baselines. Therefore, the size of the spot corresponding to the A star more or less defines the point spread function (PSF) of the image, about 1\,mas large. Because the M giant is the most luminous component of the system in the H-band, we identify it in the image as the darkest spot. With respect to the A star, we clearly see that it is spatially resolved and measures approximately 2\,mas in diameter. We expect the distortion observed in the image to come from an asymmetric filling of the $(u,v)$-plane (implying a non-circular PSF on the reconstructed image) rather than to a real tidal distortion. As a matter of fact, the tidal distortion would be around $5~\sim~7$\%, i.e.\ less than seen in the image. Additionally, its orientation in the image corresponds well with the asymmetry observed in the corresponding $(u,v)$-planes. It was actually not possible to image the circumbinary disc because of the lack of data with short baselines.

\section{Modelling} \label{part:model}

Our observations clearly show that SS Lep is a spatially resolved binary whose M giant is actually resolved for all observations. We built a geometrical model to determine the characteristics of the individual components. The M giant and the A star are modelled as uniform discs, and the circumbinary material is modelled as a Gaussian envelope. We tried to detect a possible tidal distortion of the giant or matter escaping from its atmosphere by modelling it with an elongated uniform disc. Results were not persuasive and, similar to the image reconstruction, we cannot conclude anything about this because we lack the longest baselines able to measure distortion of a few percent. The spatial resolution of 1\,mas was also not sufficient to resolve the putative shell or an accretion disc around the A star, which agrees with the 0.5\,mas size estimated from the spectral energy distribution (SED) in \citet{verhoelst_2007}. We therefore fixed its diameter to 0.5\,mas.

The model we used to fit the interferometric data { (visibilities and closure phases)} therefore comprises six degrees of freedom: the relative flux contribution of two components of the system; the binary separation and its orientation, the size of the M giant, the size of the circumbinary envelope. Our data sets are perfectly suited to spectral analysis. To properly fit the data it appeared necessary to consider the fluxes to be wavelength-dependent.

The data and the results of our fits are presented in the Appendix. { Starting from the PIONIER images,} we are able to measure the binary separation and orientation for each dataset independently. The relative flux of the three components could be recovered between 1.6 and $2.5\,\mu$m, with a dispersion of 3\% between the different epochs. The relative flux ratios are almost constant over the H-band, which validates the ``grey'' approach used in Sect.~\ref{part:image}. We tried to measure a chromatic diameter for the giant but results were not consistent between the different epochs. Finally, despite the relatively long period between AMBER and PIONIER observations (almost two orbital periods), we note rather good consistency of results within error bars, indicating that the system is relatively stable.



We were not able to extract much information about the large circumbinary disc because of the lack of very short baselines. The disc is almost totally resolved with the smallest projected baselines of 15\,m. We were only able to roughly determine its full width at half maximum (FWHM) for only two observations out of eight (A1 and A4),  and its relative flux for 6 of them (A1, A3, P1 to P4). We measured an FWHM of $12.2\pm0.2$\,mas in agreement with the estimation of \citet{verhoelst_2007} in the near-IR. The disc has been observed in the mid-IR by \citet{verhoelst_2011}, who measured a Gaussian FWHM of 26\,mas. Obervations with the 10-metre baselines of the Keck Segment-Tilting Experiment at $10.7\,\mu$m did not resolve it \citep{monnier_2009} and indicate that it should not be larger than 200\,mas in the mid-IR.


\section{Characteristics of the individual components} \label{part:binary}

%
\begin{figure}
	\centering
	\includegraphics[width=.33\textwidth]{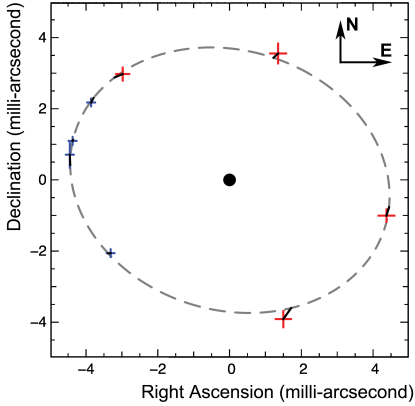}
	\caption{SS Lep best orbit (dashed line) obtained by combining previous radial velocities \citep{welty_1995} with our astrometric measurements. The central dot indicates the A star. AMBER and PIONIER points are respectively presented by the red and blue crosses representing the 3-$\sigma$ error bars. The corresponding points on the best orbit are indicated by the short segments originating in each point. 
	\label{fig:orbit}}
\end{figure}
%

\subsection{The orbit of SS Lep}

To compute the most reliable orbit possible, we combined our eight astrometric positions of the binary with the radial velocities { of the M star obtained by} \citet{welty_1995}\footnote{ We discarded the A star radial velocitites that were not convincing.}. { We deduced the orbital parameters from a global $\chi^2$ minimisation of these data}. The best-fit orbit is shown in Fig. \ref{fig:orbit}, and the orbital parameters are listed in Table \ref{tab:orbit}. Uncertainties on the orbital elements are estimated via Monte-Carlo simulations. The inclination angle of $143.7^\circ$ and the non-significant eccentricity\footnote{The astrometric points alone lead to an eccentricity of $0.004\pm0.008$, compatible with a circular orbit.} confirm the previous measurements of \citet{welty_1995} of a quasi-circularised orbit observed with an inclination angle between $28^\circ$ and $38^\circ$.
The circularised orbit is not a surprise for an evolved symbiotic system, with an M giant in a short orbit. 
For instance, \citet{fekel_2007} find that 17 of the 21 (i.e.\ 81\%)  red symbiotic systems with periods $P\leq800$ days have circular orbits.

This result definitely invalidates the periastron-passage mass transfer scenario of \citet{cowley_1967}, which required a significant eccentricity ($e=0.134$) to explain the regular ``outbursts'' of the system. Finally, given the almost null eccentricity, $\omega$ is also poorly constrained.

%
\begin{table}
	\caption{Orbital parameters of SS Lep obtained by combining previous radial velocities \citep{welty_1995} with our 8 astrometric measurements.
	\label{tab:orbit}}
\begin{centering}
	\begin{tabular}{ll}
	\hline \hline
 Semi major axis $a$  &$4.492\pm0.014$\,mas  \\
Linear semi major axis $a$   & $1.26\pm0.06\,$AU \\
 Inclination $i$& $143.7\pm0.5^\circ$ \\
 Eccentricity $e$& $0.005\pm0.003$ \\
 Longitude of the ascending node $\Omega$  &  $162.2\pm0.7^\circ$ \\
 Argument of periastron $\omega$ & $118\pm30^\circ$  \\
   	\hline
	\end{tabular}
\end{centering}
\end{table}
%

\subsection{The masses} \label{part:masses}

Combining our value for the inclination with the binary mass function obtained by \citet{welty_1995}, we can estimate the individual mass of the stars, hence the mass ratio. The main source of uncertainty in this estimation resides in the distance, as determined by Hipparcos. Using the distance and the angular separation of the two stars, we obtain $a = 1.26 \pm0.06$\,AU, and thus, through Kepler's third law, the total mass of the system is estimated as $4.01\pm 0.60  \,M_\odot$. We then derive $M_A = 2.71\pm0.27\,M_\odot$, $M_M=1.30\pm0.33\,M_\odot$, and $1/q = M_A/M_M = 2.17\pm0.35$. 
The mass ratio is thus much greater than previously thought. While the A star still has a mass in the range $2-3\,M_\odot$, the mass of the M giant is now much higher than estimated earlier, 
and this implies that less matter was transferred into the system than previously guessed. We come back to this later.

%
%

\subsection{The M star}

Averaging over all the epochs, we measure an apparent diameter for the M star $\theta_{M, UD} = 2.208\pm0.012$\,mas, where the error is computed from the dispersion of the eight estimations. The error bars do not include systematic effects -- e.g.~due to any tidal distortions --, and could be a few percent. 
%
It was also not possible to identify a dependence in the giant size as a function of the wavelength.
The previous VINCI observations of \citeauthor{verhoelst_2007} led to a higher value of $2.94\pm0.3$\,mas, most likely because theirs was the result of a one-year survey of the source, without any phase information in the interferometric data. This involved modelling the system as a symmetric object, so that it was impossible to disentangle the signatures of the rotating binary from the resolved giant one.

The conversion factor from the uniform disc to a limb-darkened one differs by a few percent depending on the authors\footnote{\citeauthor{verhoelst_2007} calculate a factor of 1.058. Using the results of \citet{hanbury_1974} and \citet{claret_2000}, we find a conversion factor of 1.044, and with the method of \citet{davis_2000}, it is equal to 1.030 \citep{vanbelle_2009}.}. We adopt a value of 1.04,  which leads to a limb-darkened diameter equal to $\theta_{M, LD} = 2.296\pm0.013$\,mas. These results agree with the limb-darkened diameter estimated from the SED of \citeauthor{verhoelst_2007} ($\theta_{M,LD} = 2.66\pm0.33$\,mas). Taking the uncertainty on the distance into account, the M giant radius is $R_M = 66.7\pm3.3\,{\rm R}_\odot$, which is 40\% smaller than previously obtained. This leads to a surface gravity $\log g \sim 0.9$.

%
\begin{figure}
	\centering
	\includegraphics[width=.4\textwidth]{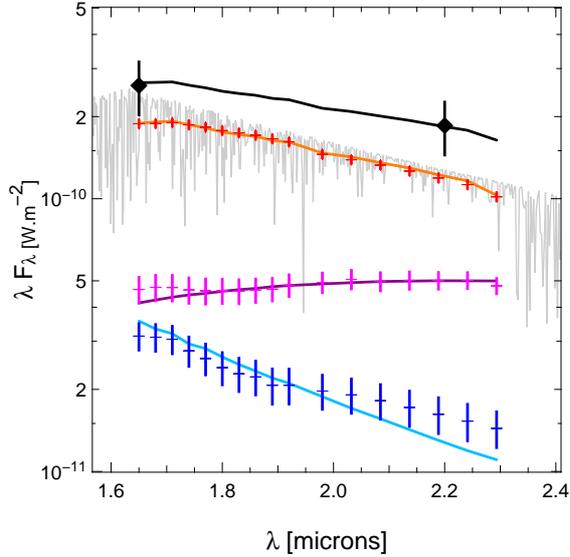}
	\caption{Flux of the M giant (red), the A star (blue), and the envelope (magenta). The grey curve is the M star MARCS spectrum. In black is the sum of the three components adjusted to the 2MASS magnitudes in the H- and K-bands. The dots are the data plus the error bars, and the solid lines are the models for each of the components.
	\label{fig:fluxes}}
\end{figure}
%

\citet{dumm_1998} provide measurements of radius for (non-Mira) M star, showing that with a radius around~$\sim 65 \: {\rm R}_\odot$, the M star spectral type should be M1/M2. Given the orbital period, \citet{muerset_1999} agree by deriving a spectral type between M0 and M1. From the table of these authors, stars with similar luminosities to the one estimated for the M giant of SS Lep all have a radius between 42 and 67 R$_\odot$.

Based on the dependence of the nuclear time scale on stellar mass, the M star must have an initial mass at least 20\% higher than its companion to have evolved on the AGB, while its companion is still in the phase of central hydrogen burning. Because the initial mass of the system was probably greater than its current value, this implies an initial mass of the giant $M_{M,0} > 2.2 \,{\rm M}_\odot$
\footnote{Since we expect the M star to have had an initial mass at least 20\% higher than the A one $q_0 = M_{M,0}/ M_{A,0}>1.2$, and therefore $M_{M,0} + M_{A,0}>1.85 M_{M,0}$. Finally, because the system was most likely more massive initially, we get  $1.85 M_{M,0} > M_A+M_M$, that is, $M_{M,0}>0.55(M_M+M_A)\simeq 2.2 \,{\rm M}_\odot$.}.
Finally, stars with such masses (unless much larger) never go through a stage with large radii when on the RGB -- with the maximum radius reached of the order of $30\,{\rm R}_\odot$ -- which indicates the M star is more likely on the early-AGB phase.

\subsection{Temperature of individual components} \label{part:fluxes}

The modelling presented in Sect.~\ref{part:model} allows us to estimate the relative flux of the three components between 1.6 and $2.3\,\mu $m (see Appendix). These measurements can be used to constrain the temperature and the size of the individual components. We impose that the M giant diameter tiso 2.2\,mas as estimated previously. We model its SED with a MARCS model whose dependencies are the temperature and the metallicity \citep{gustafsson_2008}. The A star  is in the Rayleigh-Jeans regime: it is impossible to fit simultaneously its temperature and size. We leave its diameter free and model its SED by a KURUCZ model at 9000\,K \citep{castelli_2004}. The disc is modelled by a Gaussian whose FWHM and blackbody temperature are left free. We use the absorption law of \citet{cardelli_1989} with $R_V=3.1$ and $A_V = 0.7$ mag \citep{verhoelst_2007, malfait_1998}. Additionally, we force the total SED (M giant + A star + disc) to be compatible with the 2MASS measurements \citep{skrutskie_2006}. The best fit is presented in Fig.~\ref{fig:fluxes}. The lack of an absolute spectrum makes difficult a more realistic modelling. 

The M star temperature is found to be around $3500\pm200$\,K. We also confirm that the A star is {\it apparently} larger than expected from its spectral class ($\theta_A = 0.6\pm0.05$\,mas, or a linear radius of about $18 \,{\rm R}_\odot$).  For the disc, we found a blackbody temperature of $1700\pm100$\,K and an FWHM of $8.0\pm0.5$\,mas. Interestingly, this is incompatible with the $12$\,mas derived from the fit of visibility curves in Sect.~\ref{part:model}. We see this inconsistency as a hint that a Gaussian geometry is probably too simple to model the circumbinary environment.


%
\begin{table}
	\caption{Stellar parameters extracted in this study. \label{tab:stellar_param}}
	\centering
	\begin{tabular}{lcc}
	\hline \hline
	& M star & A star \\
\hline
Mass [${\rm M}_\odot$]	& $1.30\pm0.33$ & $2.71\pm0.27$ \\
Apparent diameter [mas] & $2.208\pm0.012$ & $0.6\pm0.05$\\
Linear radius [${\rm R}_\odot$] & $66.7\pm3.3$ & $\sim18$\\
Temperature [K] & $3500\pm200$ & $\sim9000$ \\
\hline
	\end{tabular}
\end{table}
%

\section{The mass transfer process} \label{part:masstransfer}

As explained above, SS~Lep shows evidence of mass transfer between the M giant and the A star, and this mass transfer is not completely conservative. We now revisit the possible physical foundations for this mass transfer according to the new parameters of the system derived in previous sections.

The observations suggest that the mass transfer is driven by a wind-Roche Lobe overflow  \citep{podsiakolski_2007}. We indeed show here that the current state of the system seems to require an enhanced mass loss from the giant and that this wind possibly fills the Roche lobe and makes the mass transfer almost conservative. We also show that it is quite possible that an accretion disc formed around the A star, which may explain its abnormal luminosity.

\subsection{Mass transfer by Roche Lobe overflow from the wind} \label{part:rochelobe}

\begin{figure}
	\centering
	\includegraphics[width=.3\textwidth]{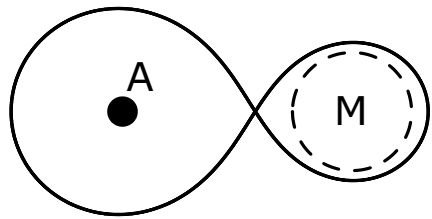}
	\caption{Representation of the modified Roche equipotential (solid line) for a mass ratio $1/q = 2.2$. The limb-darkened diameter of the M giant is the dashed line, while the A star one is the dark dot.  \label{fig:RocheEquipotential}}
\end{figure}

Our results indicate that the M giant only fills around $85\pm3$\% of its Roche lobe (Fig. \ref{fig:RocheEquipotential}). This contradicts the results of \citet{verhoelst_2007} . The reasons of this difference are threefold.
\begin{enumerate}
\item Our more precise interferometric measurements led us to estimate a smaller giant radius than in previous studies. 
\item The revised HIPPARCOS distance brings the system closer than previously thought. This makes the stars smaller, while the orbital radius is mostly given by the orbital period and is almost independent of the distance.
\item We have determined for the first time the mass ratio and find a higher value than previously guessed. This leads to a higher mass and Roche lobe radius of the giant.
\end{enumerate}
Our results disprove a {\it stricto sensus} current RLOF. Two mechanisms might occur that leave this possibility open however, but they can be discarded with quantitative arguments.

\smallskip \noindent
First, {\it radiation pressure} reduces the gravitational acceleration influence, so that the Roche potential surface shrinks \citep{schuerman_1972}. \citet{dermine_2009} estimated the ratio of the radiation to the gravitational force to be $f \sim 10^{-2}$ to $10^{-1}$ for the M giant of SS Lep. It reduces the Roche lobe radius by only 1 to 4\%. For the giant to fill its Roche lobe we need $f=0.35$, which implies that the $L_1$ and $L_2$ points share the same equipotentials, making it more difficult for the A star to accrete, most of the matter finally going into the circumbinary disc. However, the mass of the latter is rather low according to \citet{jura_2001} ($M_{\rm dust}\sim 2\times10^{-5} M_\odot$), so that this is unlikely.

\smallskip \noindent
Second, {\it atmosphere stratification} in red giants indicates that there is no single radius value, and it may be not obvious what really fills the Roche lobe. \citet{pasteter_1989} have shown that for very evolved stars the scale height of the density stratification in their atmosphere is a significant fraction of their photospheric radius, so that mass flows through the inner Lagrangian point $L_1$ long before the photosphere reaches the critical Roche equipotential. Because it is on the early AGB phase, the M giant in SS Lep is still very far from these evolutionary stages, so that this effect is negligible here.

\bigskip
Finally, to discard the RLOF mechanism, we should explain the 0.0126 mag ellipsoidal variablility in the visible from \citet{koen_2002}. This variability can be interpreted as the signature of a distorted photosphere, whose flux in the line of sight varies periodically. In the present case, it would correspond to a change in radius of about 8-9 \%. However, given the low inclination of the system, even if the star was filling its Roche lobe, it would not show such a large variation in radius. Moreover, the data in the K-band of \citet{kamath_1979}, where the M giant dominates, are hardly convincing evidence of any periodic variations in SS Lep. The Hipparcos variability must thus have another origin than a tidal distortion of the M giant. If the reason for the light variation in the visible is the primary source of light in the system, a change of less than $0.6\%$ is required to explain the observed amplitude. The cause of this change is, however, not known.

We therefore conclude that an RLOF is unlikely in SS Lep. \citet{podsiakolski_2007} suggest the possibility of a new mode of mass transfer -- the wind Roche lobe overflow -- where a slow wind fills the Roche lobe (e.g. Mira stars in symbiotic systems). Because the wind speed in M giants is rather small (around $10\sim15$\,km\,s$^{-1}$) and lower than the orbital one for SS Lep ($v_{\rm orb}=48$\,km\,s$^{-1}$), we expect it to be in the particular case of a wind Roche lobe overflow, where a substantial part of the stellar wind can be accreted. The simulations of \citet{nagae_2004} show that at least 10\% of the M giant wind could be accreted in SS Lep.

\subsection{Need for enhanced mass loss from the M giant}\label{part:transfer}

As seen in the previous section, the mass transfer in SS Lep is likely due to high-efficiency accretion of the M giant stellar wind. However, it is difficult to explain the current system state with normal stellar wind rates, which are too low. Indeed, before the AGB phase, the typical mass loss rates are around $\sim$1-$2\times10^{-8}\,{\rm M}_\odot$yr$^{-1}$ at the normal (e.g. Reimers) rate. The M giant (with an expected initial mass $> 2.2\,{\rm M}_\odot$) should have lost only a few hundredths of a solar mass before reaching the AGB, whereas we expect it to have lost at least $0.9 \,{\rm M}_\odot$. As the M star is on the AGB since only a few million years -- and will stay there for a few million years more at most -- it cannot have lost much mass since then.

There is, however, some evidence of enhanced wind mass loss of giants in binaries compared to single giants of the same spectral type \citep{mikolajewska_2007,jorissen_2003}. From a theoretical point of view, the presence of a companion reduces the effective gravity of the mass-losing star, thus enhancing the mass loss. In the case of SS Lep, the so-called CRAP mechanism of tidally enhanced stellar wind \citep{tout_1988} allows a mass loss rate 150 times higher than the Reimers rate\footnote{Because the M star fills more than 50\% of its Roche lobe, we consider an enhancement factor $1+B/2^6=156$, where $B\sim10^4$, as described by \citet{tout_1988}.}, i.e.~$\sim 2.4\times10^{-6} \,{\rm M}_\odot$yr$^{-1}$. \citet{soker_1998} and \citet{frankowski_2001} have also shown that, in this case, the mass loss is strongly enhanced in the equatorial plane, while an accretion disc can form during wind accretion \citep{theuns_1996, nagae_2004}.

To validate our scenario of a wind RLOF and of enhanced mass loss by wind from the giant, we have considered the possible evolution of a binary system, taking the CRAP mechanism into account, and following the methodology of \citet{hurley_2002}.
We start with a system having an initial period of 160 days and initial masses $M_{\rm M,0}=2.28  \,{\rm M}_\odot, \: M_{\rm A,0}=1.85  \,{\rm M}_\odot$. For about 1 Gyr, the system evolves without much change, and the primary star starts its ascent of the AGB. After 2.8 Myr, the masses and period have reached the currently observed values, with about $0.1 \,{\rm M}_\odot$ having been lost by the system, and forming some circumbinary disc. The mass loss and transfer occurred mostly during the last 500 000 years, with a mass loss $\sim~10^{-6}\,{\rm M}_\odot$yr$^{-1}$. During the whole process, the Roche lobe radius around the M star remained similar, the lowest value being $74\,{\rm R}_\odot$. No RLOF should thus have happened, unless the initial eccentricity was very high.
%

The above scenario is certainly not the only possible one, but it shows that we can explain the current and peculiar properties of SS Lep (including a low-mass circumbinary disc), without resorting to an RLOF and with a low mass loss rate of the order of $10^{-6}\,{\rm M}_\odot$yr$^{-1}$.

\subsection{Accretion on the A star} 

From the previous sections, the accretion efficiency is expected to be much higher than 10\% in SS Lep, perhaps in the rage of 80-90\%, while the mass loss rate could reach around $10^{-6}\,{\rm M}_\odot$yr$^{-1}$. This corresponds to an accretion rate of a few $10^{-7} \sim 10^{-6}\,{\rm M}_\odot$yr$^{-1}$ on the A star. This agrees with the $3.3\sim5.5\times10^{-7}\,{\rm M}_\odot$yr$^{-1}$ of  \citet{blondel_1993}, when interpreting the Lyman-$\alpha$ emission in terms of the recombination of H$^+$ during the inflow of matter.

\citet{verhoelst_2007} propose that the A star could have ``puffed up'' by factors of five to seven because of too fast accretion in order to explain its abnormal luminosity. The calculations of \citet{kippenhahn_1977} require an accretion rate of $ 5\times10^{-5} {\rm M}_\odot$yr$^{-1}$ to reconcile the luminosity and the spectral type for an increase in diameter of a factor of ten \footnote{Based on the same computations, \citeauthor{verhoelst_2007} derived a value of $\sim 2\times 10^{-4}\,{\rm M}_\odot$yr$^{-1}$, but the calculations of \citeauthor{kippenhahn_1977} give an increase in radius by a factor of 100 for this accretion rate.}. This is too high compared to the value we derived previously. Moreover, in this scenario a shell spectrum is hiding the A star spectrum so that a clear determination of the stellar parameters is prevented; in particular, the gravity of the star cannot be estimated accurately. If we make the reasonable assumption that the A star is rotating in the orbital plane, based on the measured $v\, \sin i = 118\,$km\,s$^{-1}$ \citep{royer_2002}, we derive an equatorial velocity of 196\,km\,s$^{-1}$. It is above the break-up speed for a $2.7 \,{\rm M}_\odot$ star with a radius of $18\,{\rm R}_\odot$, which is 170\,km\,s$^{-1}$. These arguments challenge the ``puffed up'' scenario, so we can wonder if the A star indeed has such a large radius.

Another explanation for the abnormal luminosity of the A star could be the presence of an accretion disc. Indeed, the Ly-$\alpha$  emission profile observed by \citet{blondel_1993} consists of a single, asymmetric and redshifted feature, suggesting a significant absorption by the extended atmosphere in the equatorial plane. This could be a hint of wind RLOF presenting a focused wind in the equatorial plane and of the presence of an accretion disc. This disc would easily explain the shell nature of SS Lep as well as the variability of its spectrum. This accretion disc is also expected from a mechanical point of view in the wind RLOF scenario because most of the matter goes through the $L_1$ point. The matter will fall towards the A star and, because of its initial angular momentum, it will reach a minimum radius of 
$$\frac{r_{\rm min}}{a} = 0.0488\,q^{-0.464},$$
before forming an accretion disc. In the present case, this amounts to $20\,{\rm R}_\odot$, larger than the star (even if it has expanded). Moreover, at the beginning of the mass transfer, $r_{\rm min} \simeq 9 \,{\rm R}_\odot$, which is also larger than the radius of an A star on the main sequence ($\sim$\,2-3\,R$_\odot$). It is thus likely that a disc has formed. The radius of this disc would be about 
$$\frac{r_{\rm circ}}{a} = 0.0859\,q^{-0.426},$$
which is $r_{\rm circ} \simeq 33 \,{\rm R}_\odot$ currently for SS Lep. The expected apparent diameter of this disc is between 0.8 and 1\,mas if we take the inclination into account. This size is at the limit of detection with our VLTI interferometric observations but is within reach for CHARA instruments like MIRC or VEGA. We think this alternative deserves further scrutiny.

\subsection{Future evolution}

It is also interesting to consider the future of SS Lep in the framework of the model presented in Sect.\,\ref{part:transfer}. As the M star will evolve along the AGB, its mass loss will increase more and more. Given the multiplication factor imposed by the CRAP mechanism, after 170 000 years, the envelope will be exhausted and the star will start its post-AGB evolution to become a WD. It is likely that for a short time, the system will appear as a non-spherical planetary nebula, where the asymmetry is due to the circumbinary material. The A star will have its mass increased to $3.3\,{\rm M}_\odot$. Because the primary will not have time to go through the thermal pulses phase, it will most likely not be polluted in s-process elements. The period will have increased to 900 days, so the system will appear in the typical location in the ($e-\log P$) diagram for post-mass transfer systems \citep{boffin_1993}. When in the post-AGB phase, the system will thus appear typical of those discovered, with a circumbinary disc and an orbital period between 200 and 1800 days \citep{vanwinckel_2003}. The A star will then undergo its evolution as a red giant and the system will most likely again be a symbiotic system -- more usual this time, given the presence of a white dwarf.

\section{Conclusion and future work}

We have presented here the results of VLTI observations, and we focused on the binary. After having computed the characteristics of the orbit, we demonstrated that the mass ratio is lower than previously thought and that the M giant does not fill its Roche lobe. However the system is possibly in the configuration of a wind Roche lobe overflow, where a substantial part of the giant's stellar wind can be accreted by the A star. We also have good reasons to think that the A star is actually surrounded by an accretion disc, although this needs to be investigated further.

We still lack a low-resolution spectro-photometry of SS~Lep between $1.6$ and $2.5$\,$\mu$m to compute the absolute luminosity of each component and extract more specific information. The current data also present good hints of unmodelled material escaping the system. It would be of uttermost importance to obtain additional interferometric observations of SS Lep -- at large but also small baselines -- to study the binary interactions in more details (M star distortion, mass transfer, circumbinary disc) and to determine if the A star is surrounded by an accretion disc.

\acknowledgement{The authors would like to thank O.\ Absil and G.\ Dubus for their help, as well as the referee N.\ Elias, whose careful review of the text helped them improve the papers clarity and quality. PIONIER is funded by the Universit\'e Joseph Fourier (UJF, Grenoble) with the programme TUNES-SMING, the Institut de Plan\'etologie et d'Astrophysique de Grenoble (IPAG, ex-LAOG), and the Institut National des Science de l'Univers (INSU) with the programmes ``Programme National de Physique Stellaire'' and ``Programme National de Plan\'etologie''. PIONIER is equipped with a detector provided by W. Traub (JPL, Caltech). The authors want to warmly thank the VLTI team. This work is based on observations made with the ESO telescopes. It made use of the Smithsonian/NASA's Astrophysics Data System (ADS) and of the Centre de Donnees astronomiques de Strasbourg (CDS). All calculations and figures were performed with the freeware \texttt{Yorick}\footnote{\texttt{http://yorick.sourceforge.net}}. The image reconstruction was performed with the software \texttt{MIRA}\footnote{\texttt{http://www-obs.univ-lyon1.fr/labo/perso/ eric.thiebaut/mira.html}}.
}


\bibliographystyle{aa}

\listofobjects

\appendix
\onecolumn

\section{Data and results of the parametric modelling} \label{app:fit_result}

\begin{figure*}[h!]
\centering
\includegraphics[width=0.22\textwidth]{UV_AMBER54781}
\includegraphics[width=0.22\textwidth]{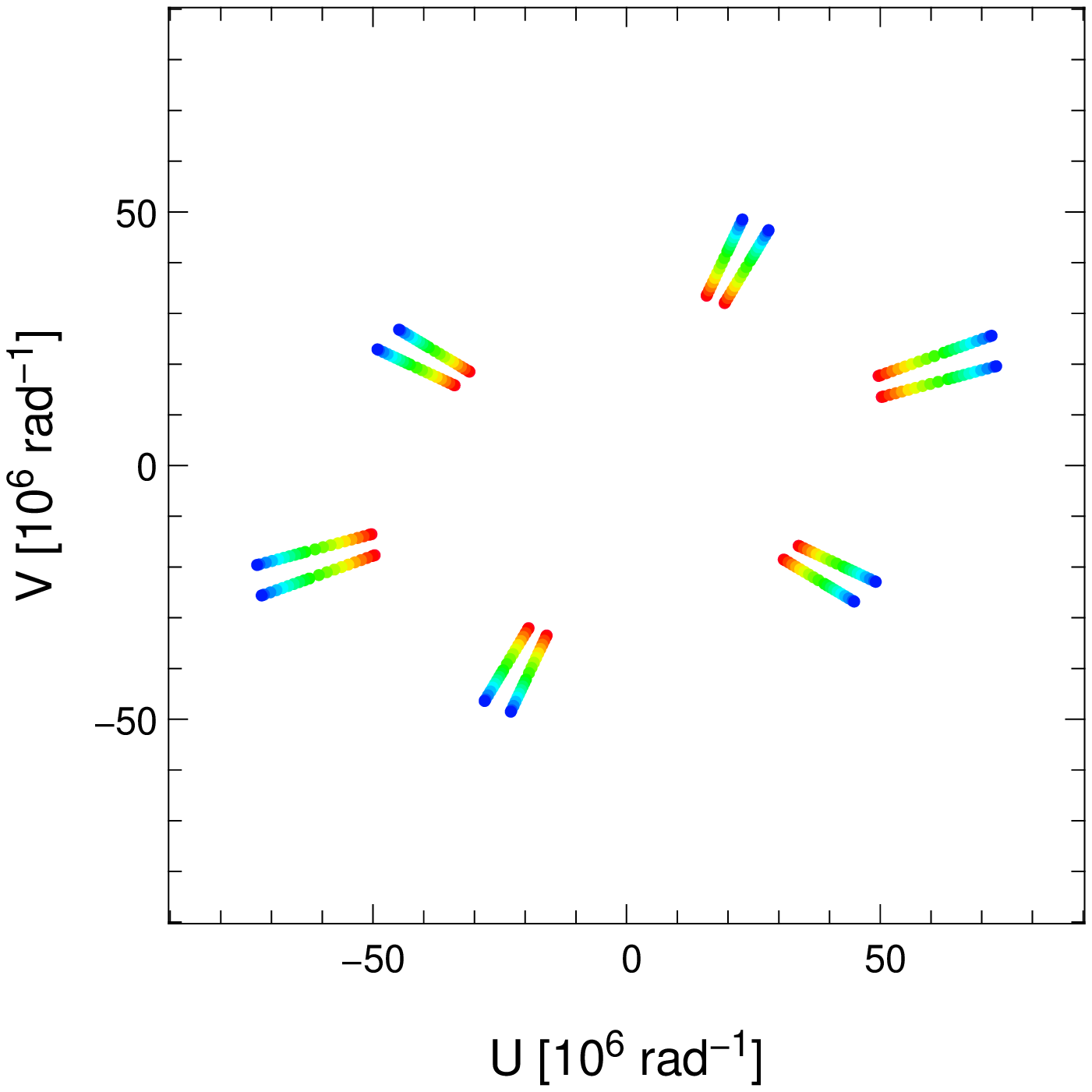}
\includegraphics[width=0.22\textwidth]{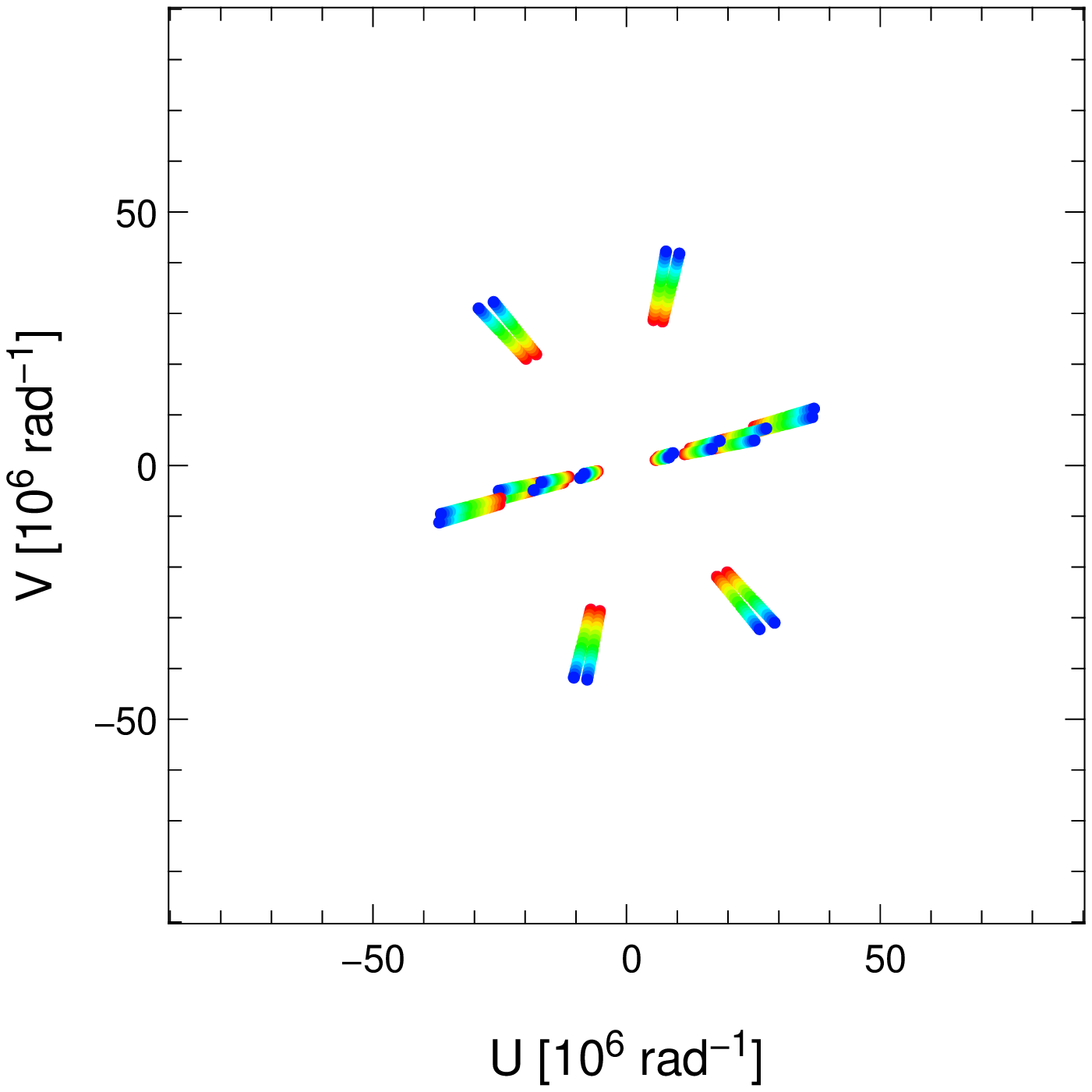}
\includegraphics[width=0.22\textwidth]{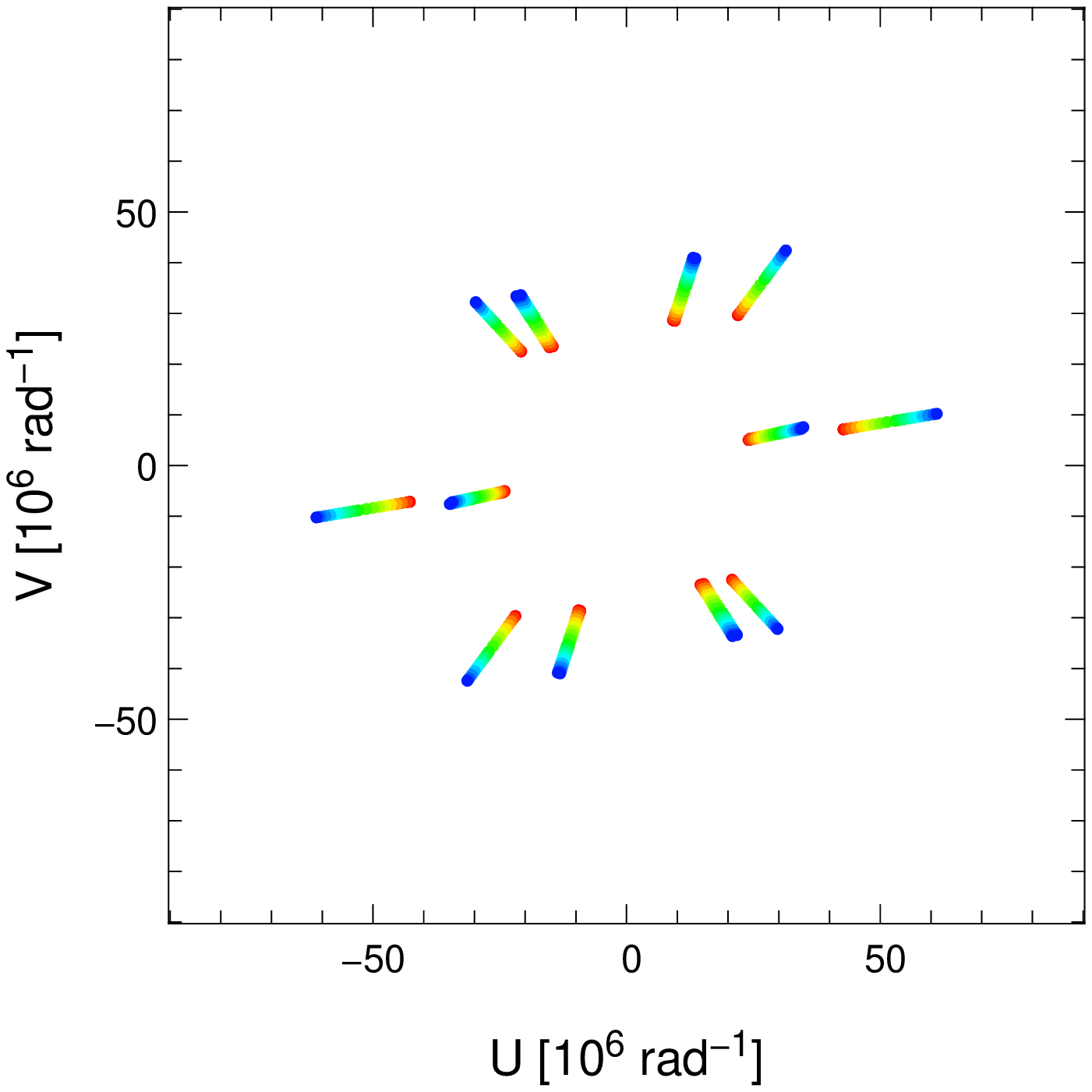}\\
\includegraphics[width=0.22\textwidth]{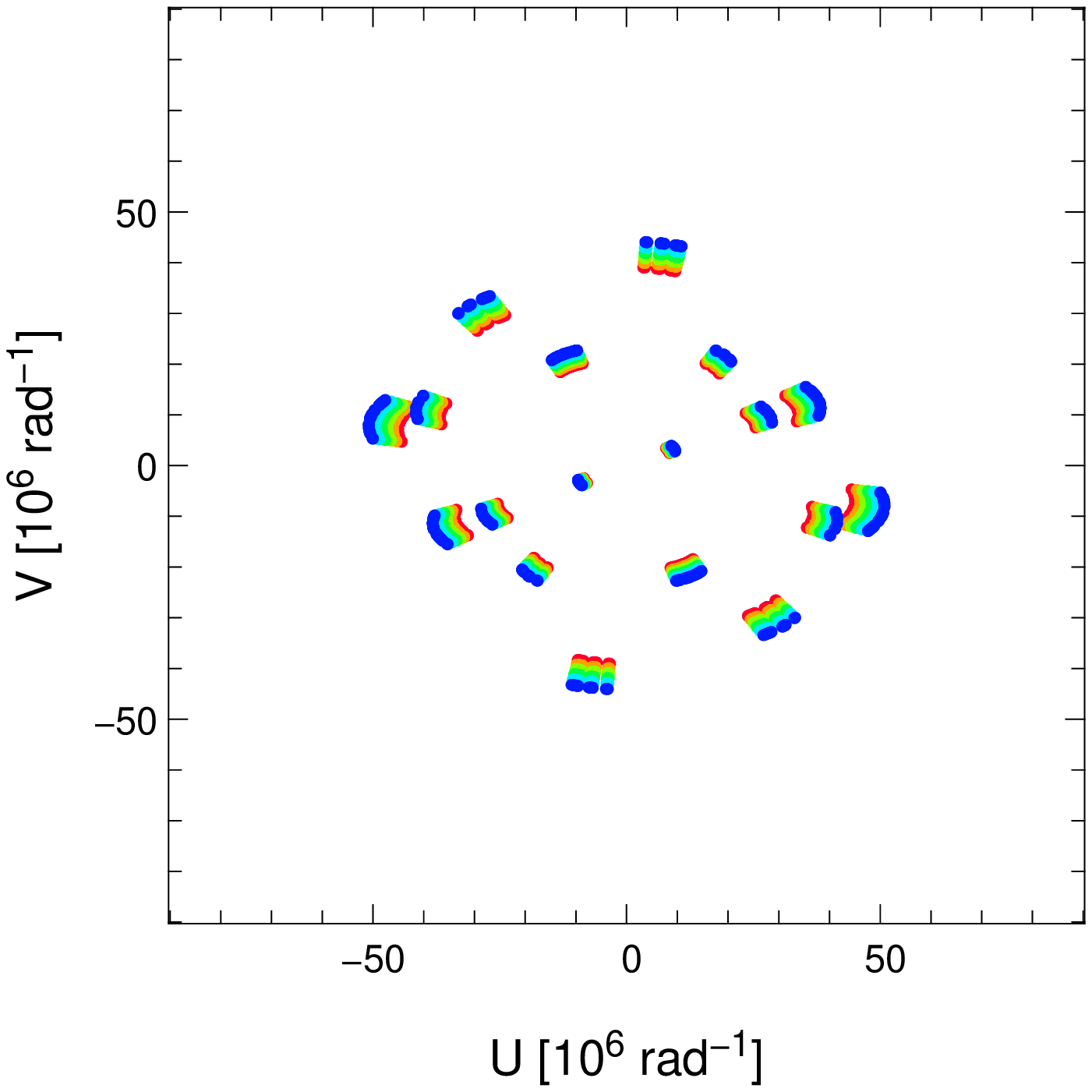}
\includegraphics[width=0.22\textwidth]{UV_PIONIER55529}
\includegraphics[width=0.22\textwidth]{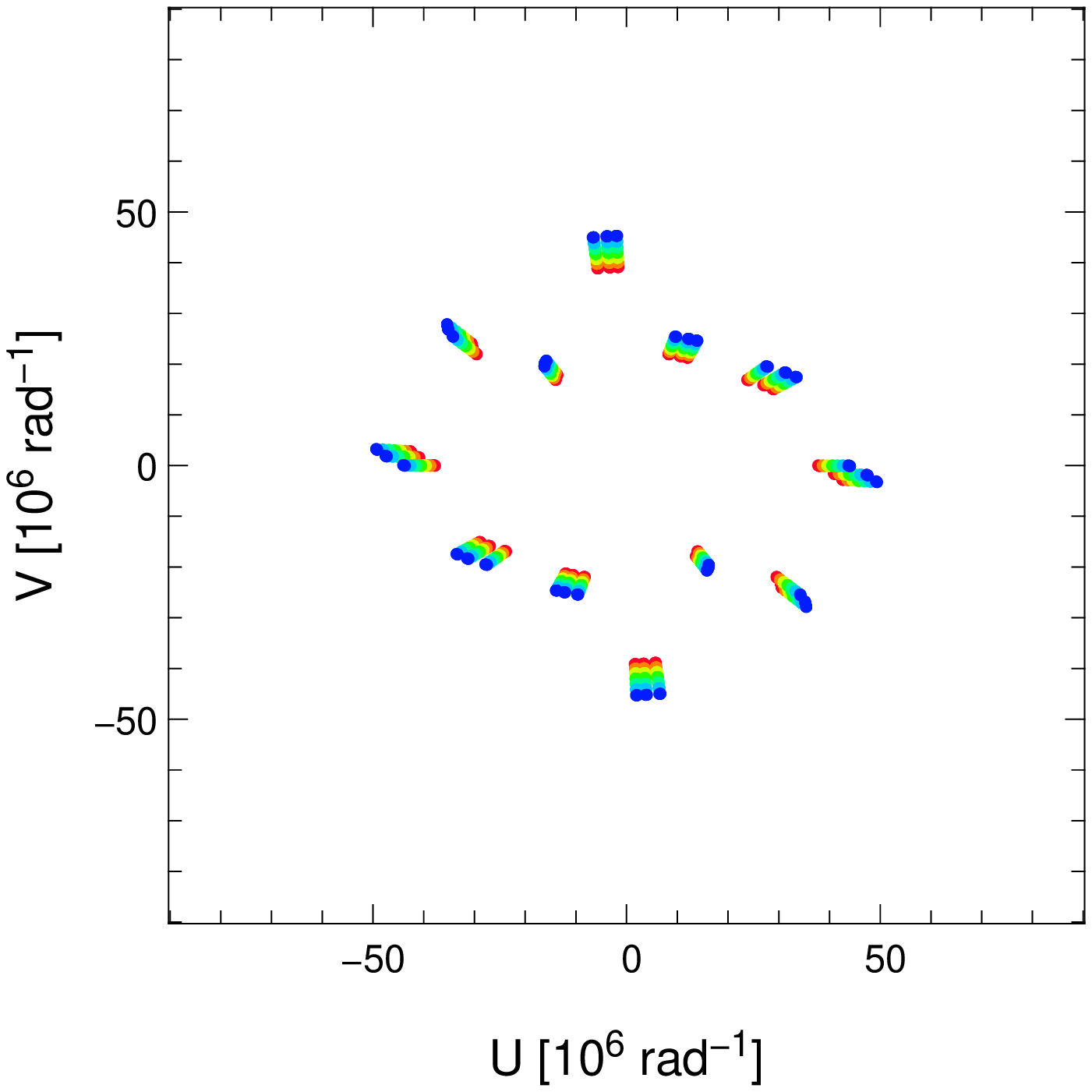}
\includegraphics[width=0.22\textwidth]{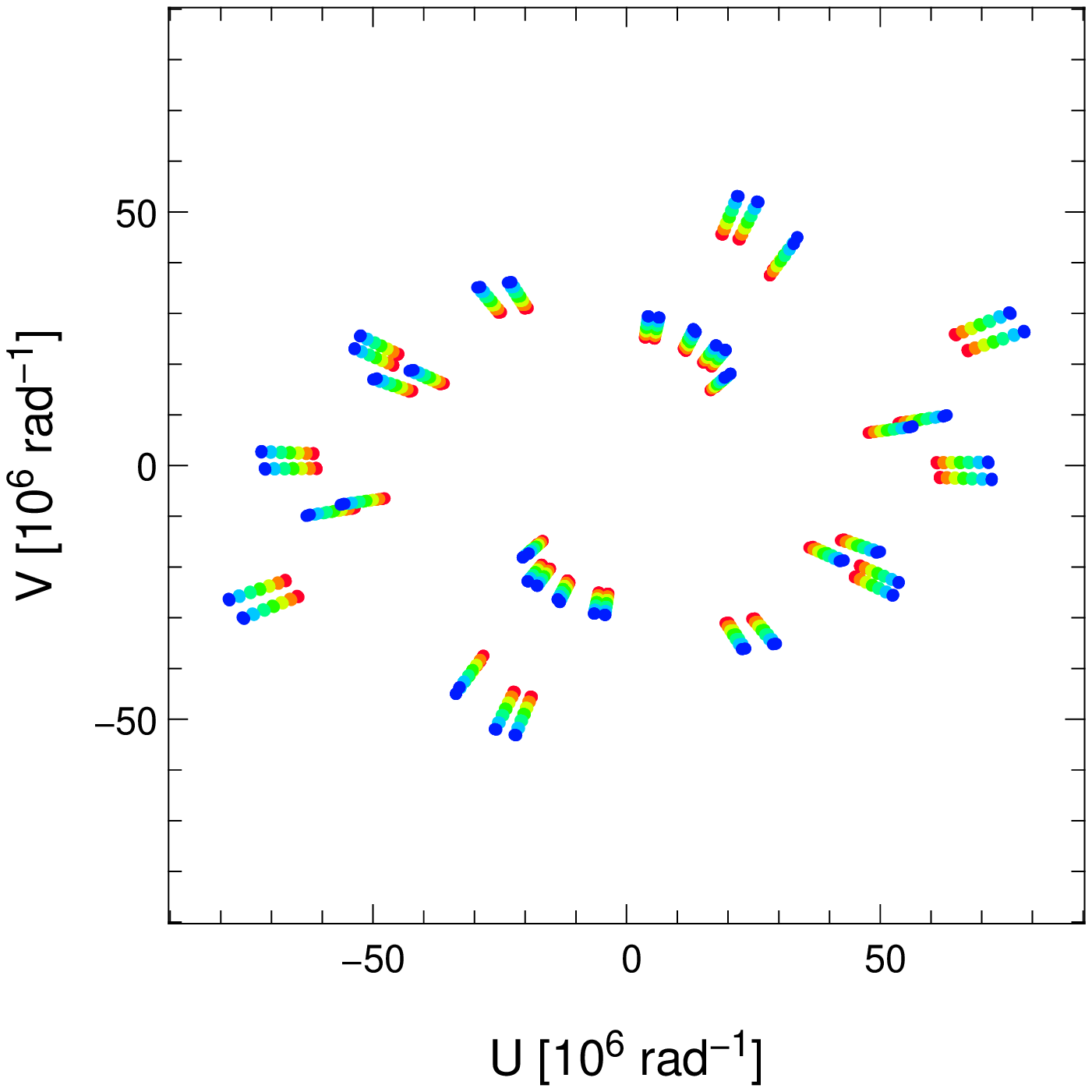}
\caption[]{{\it (u,v)-}plane coverage for observations of SS Lep (top: AMBER; bottom: PIONIER). \label{fig:UVSSLep}}
\end{figure*}

\begin{figure*}[h!]
\centering
\includegraphics[width=0.22\textwidth]{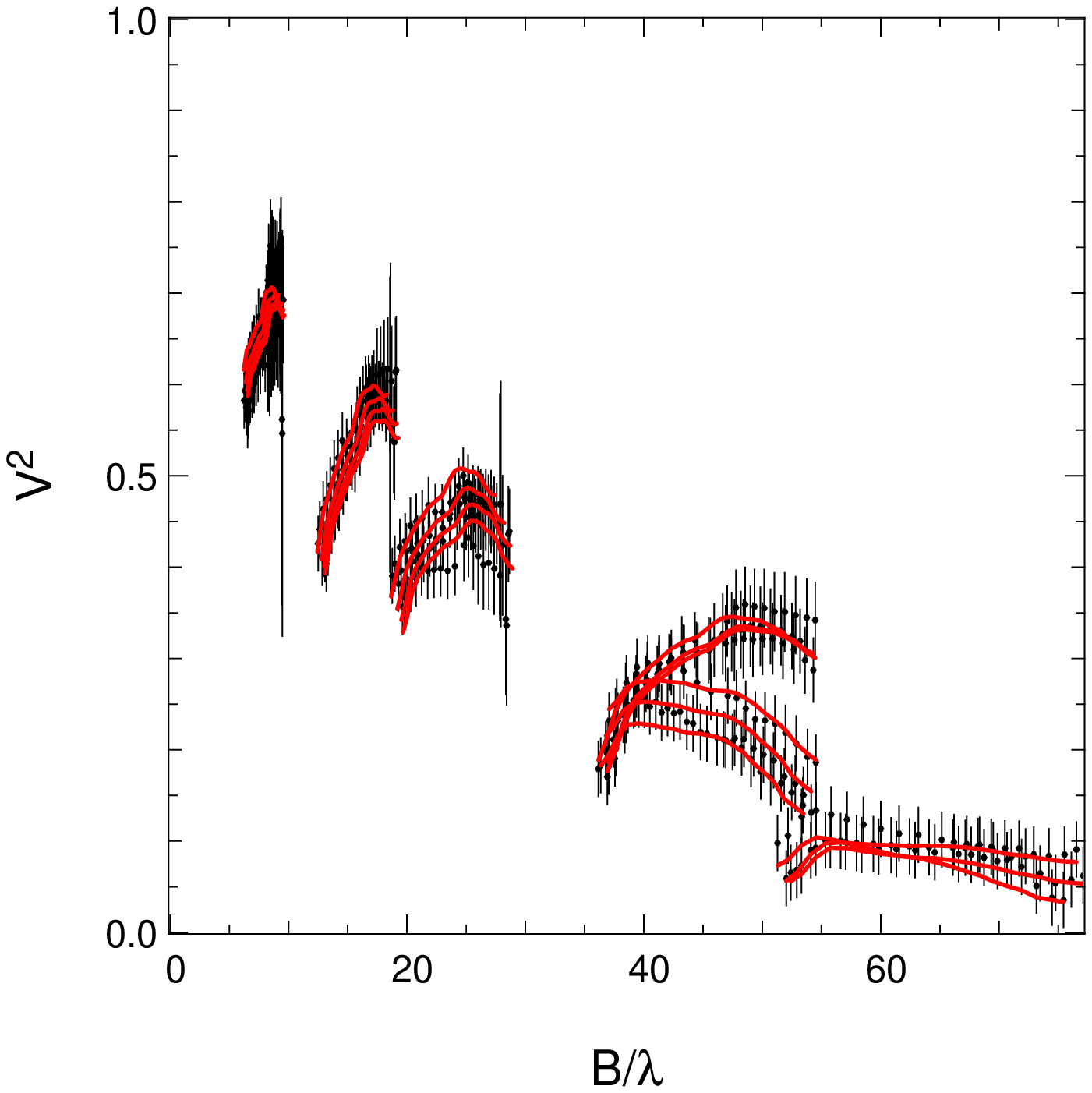}
\includegraphics[width=0.22\textwidth]{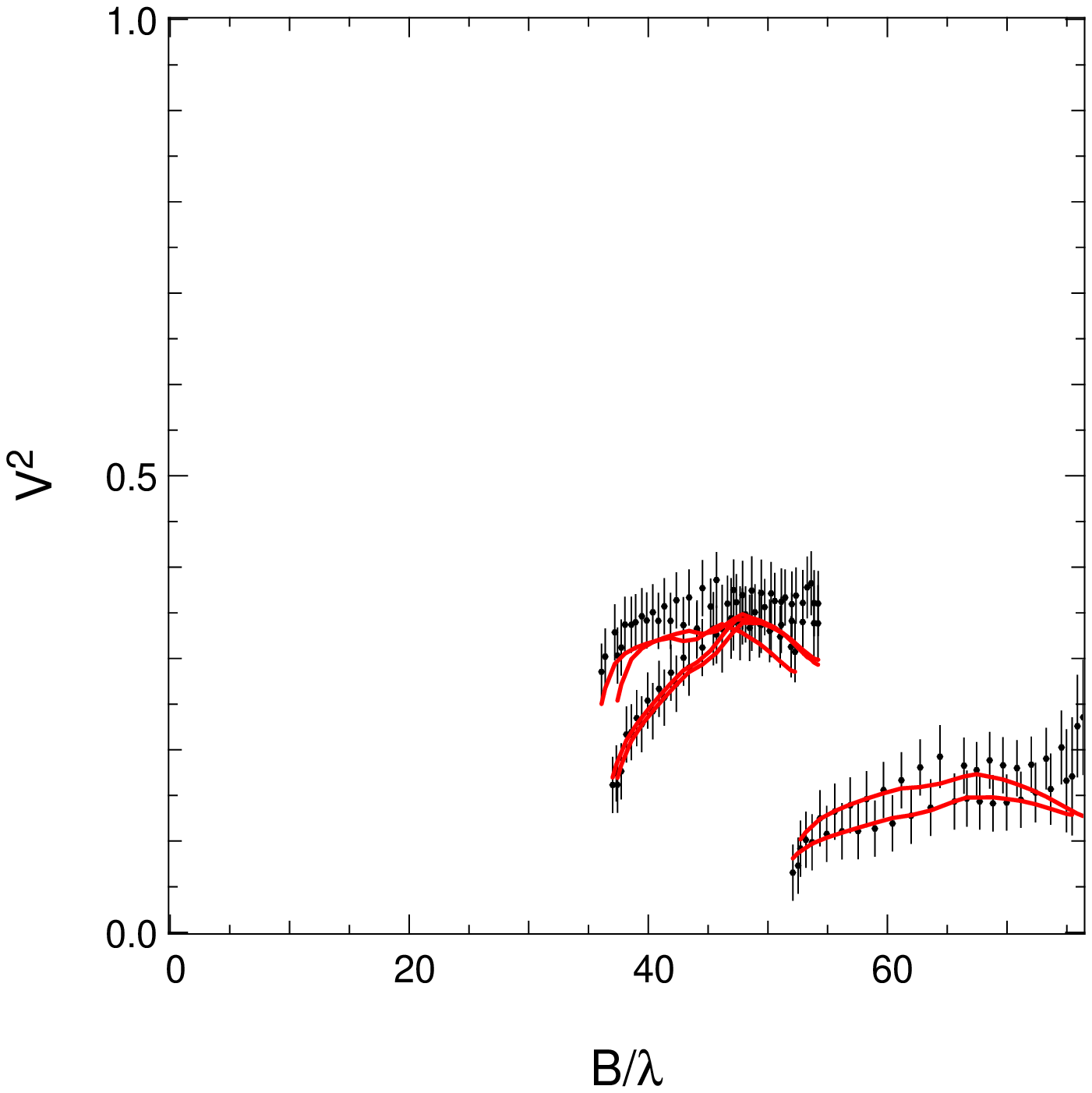}
\includegraphics[width=0.22\textwidth]{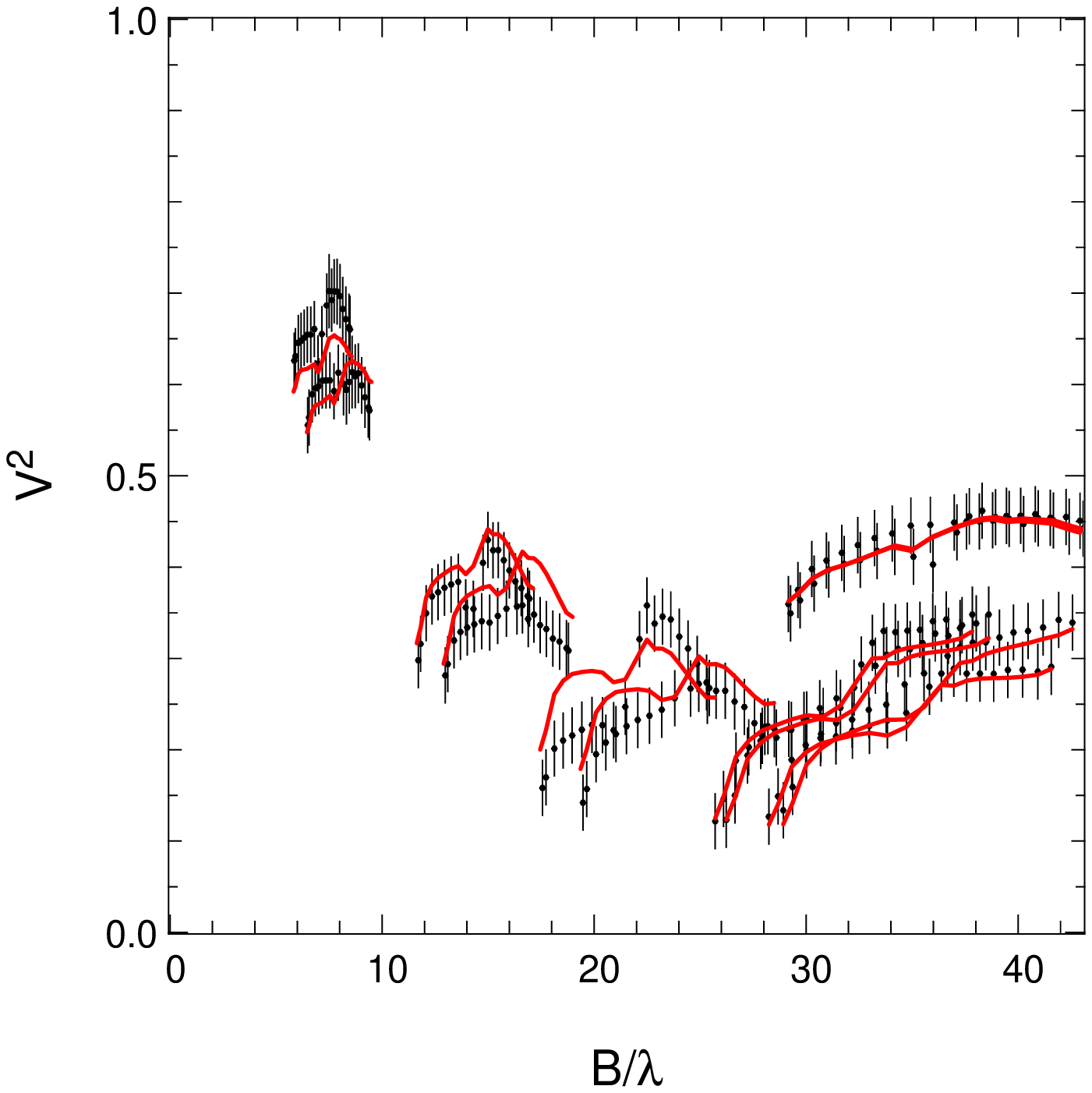}
\includegraphics[width=0.22\textwidth]{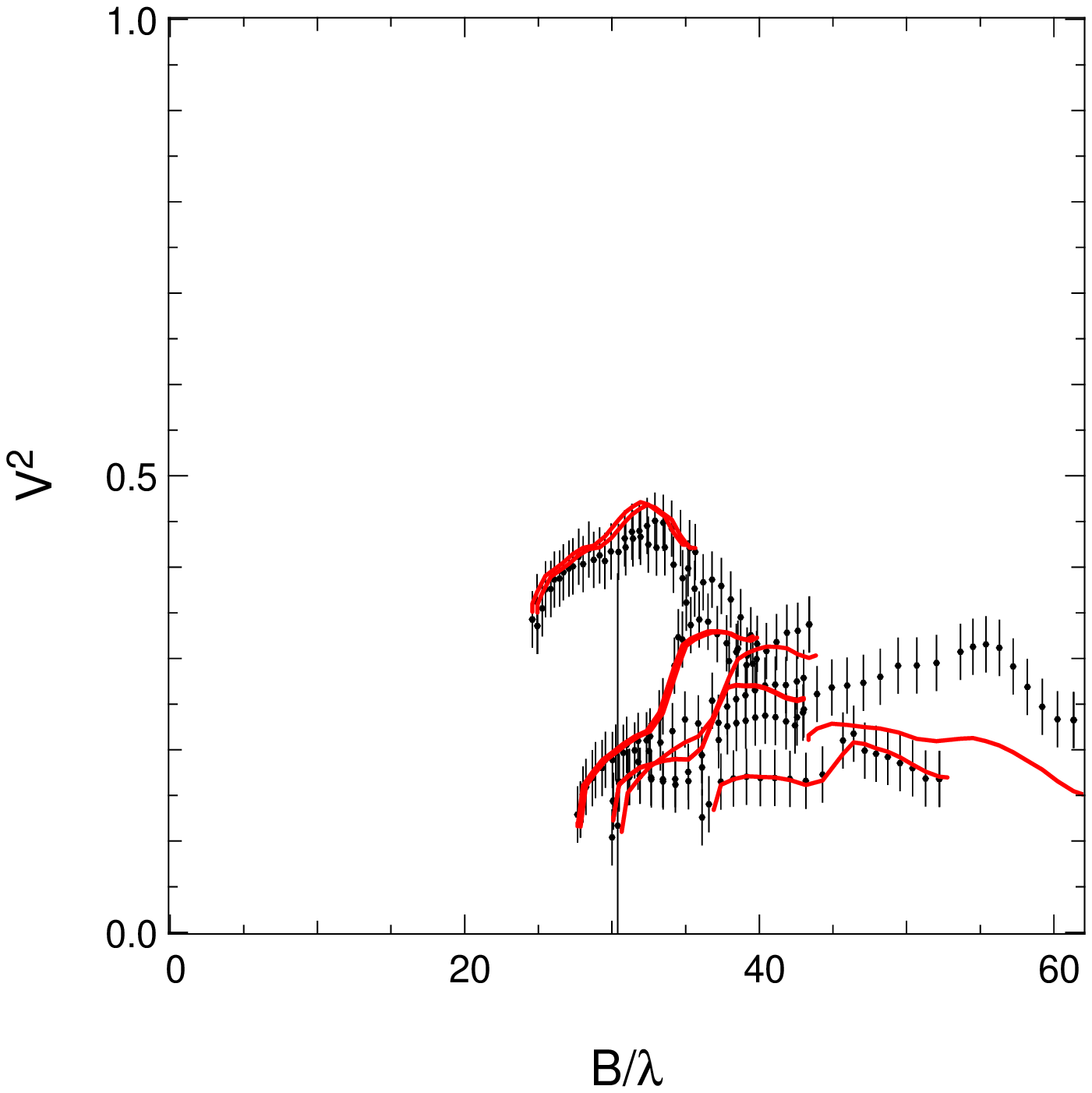}\\
\includegraphics[width=0.22\textwidth]{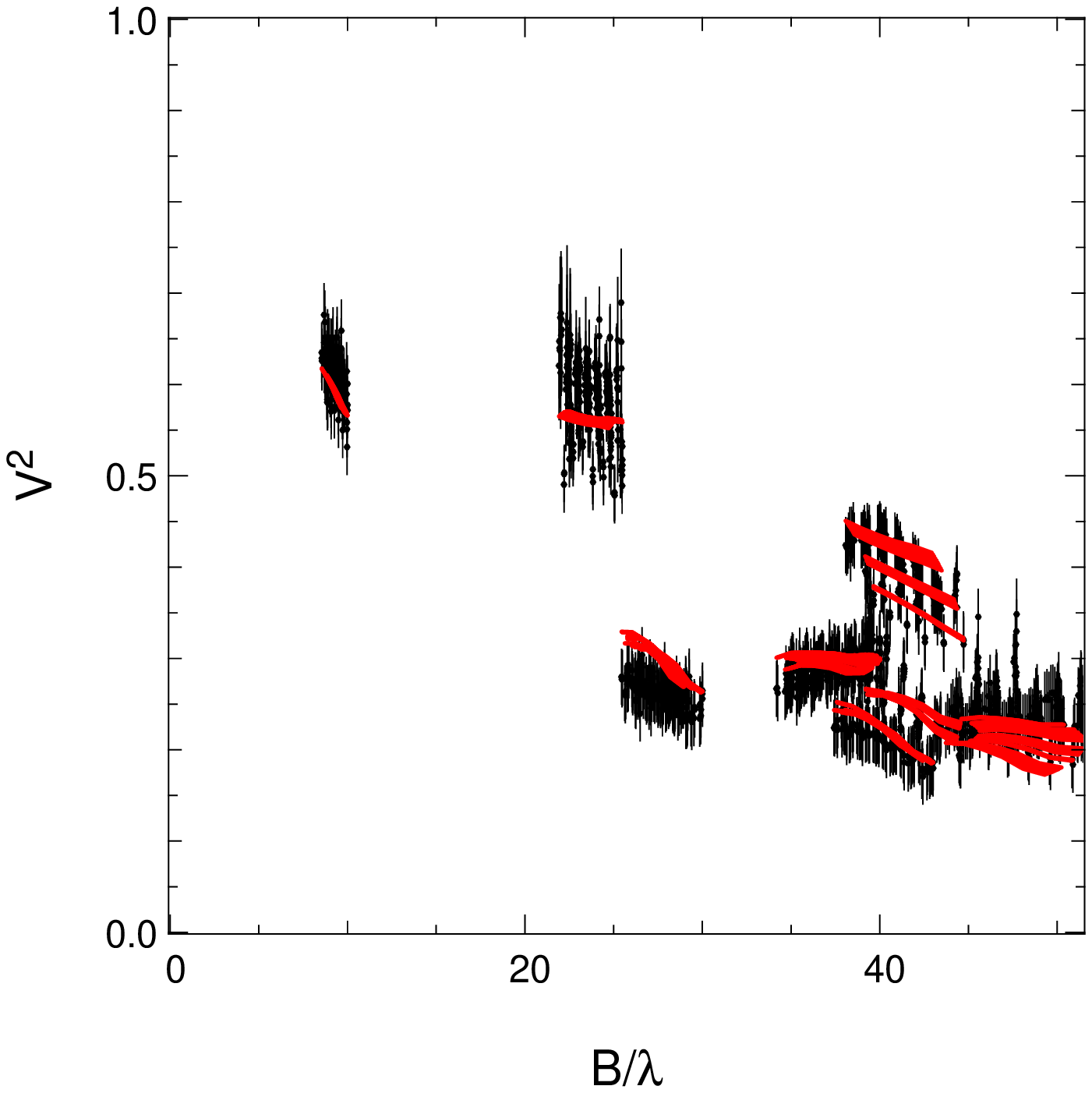}
\includegraphics[width=0.22\textwidth]{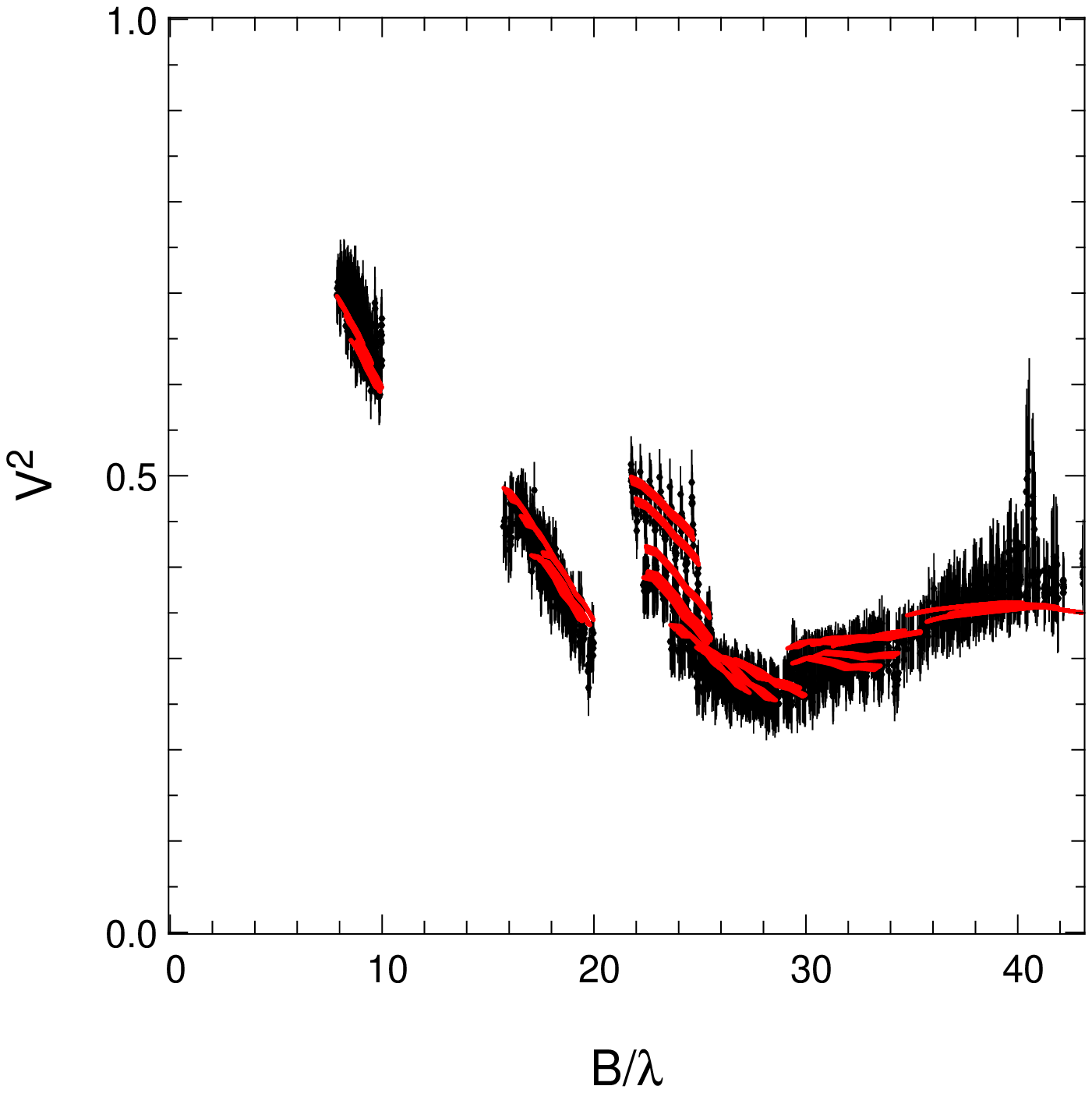}
\includegraphics[width=0.22\textwidth]{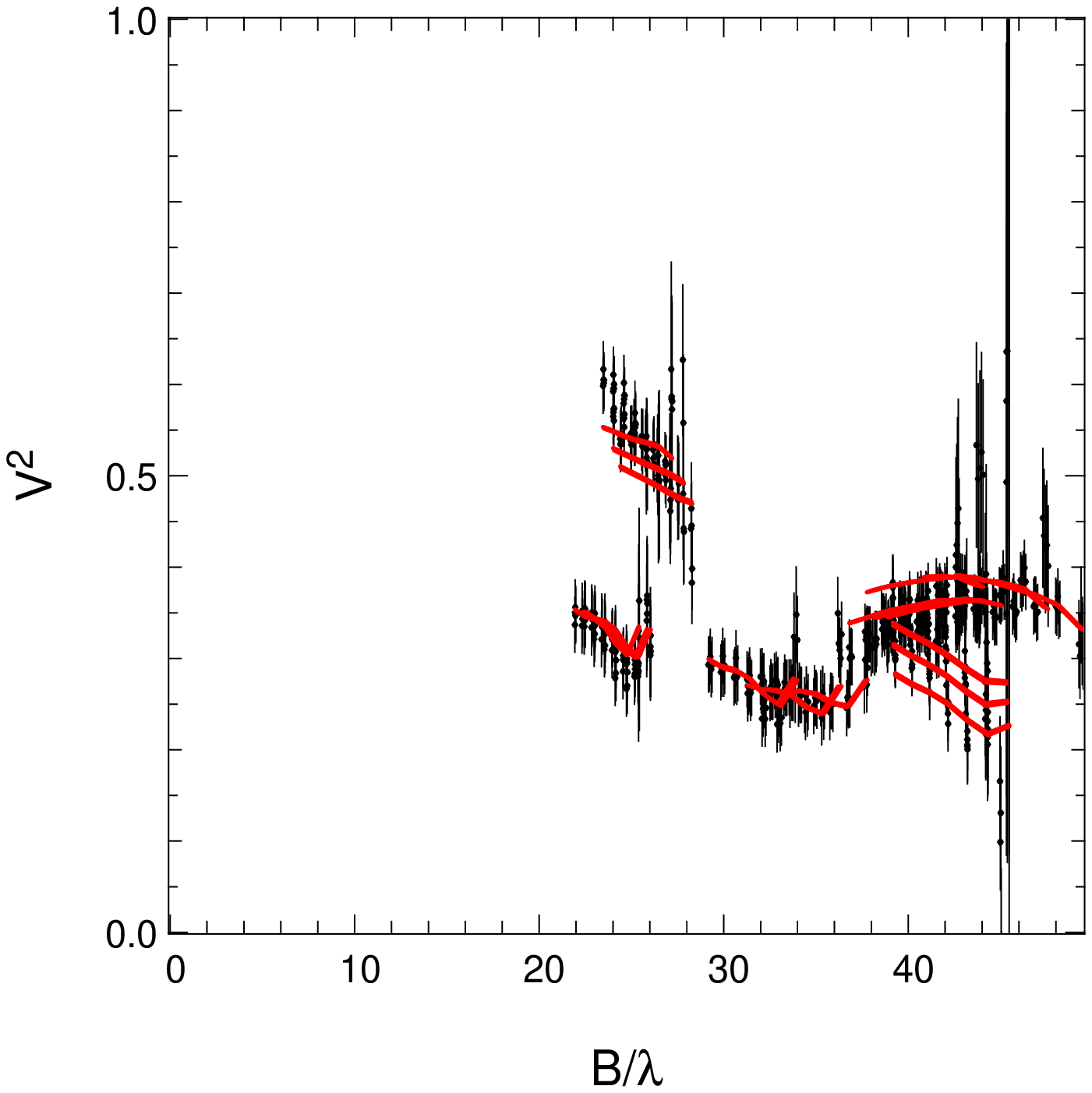}
\includegraphics[width=0.22\textwidth]{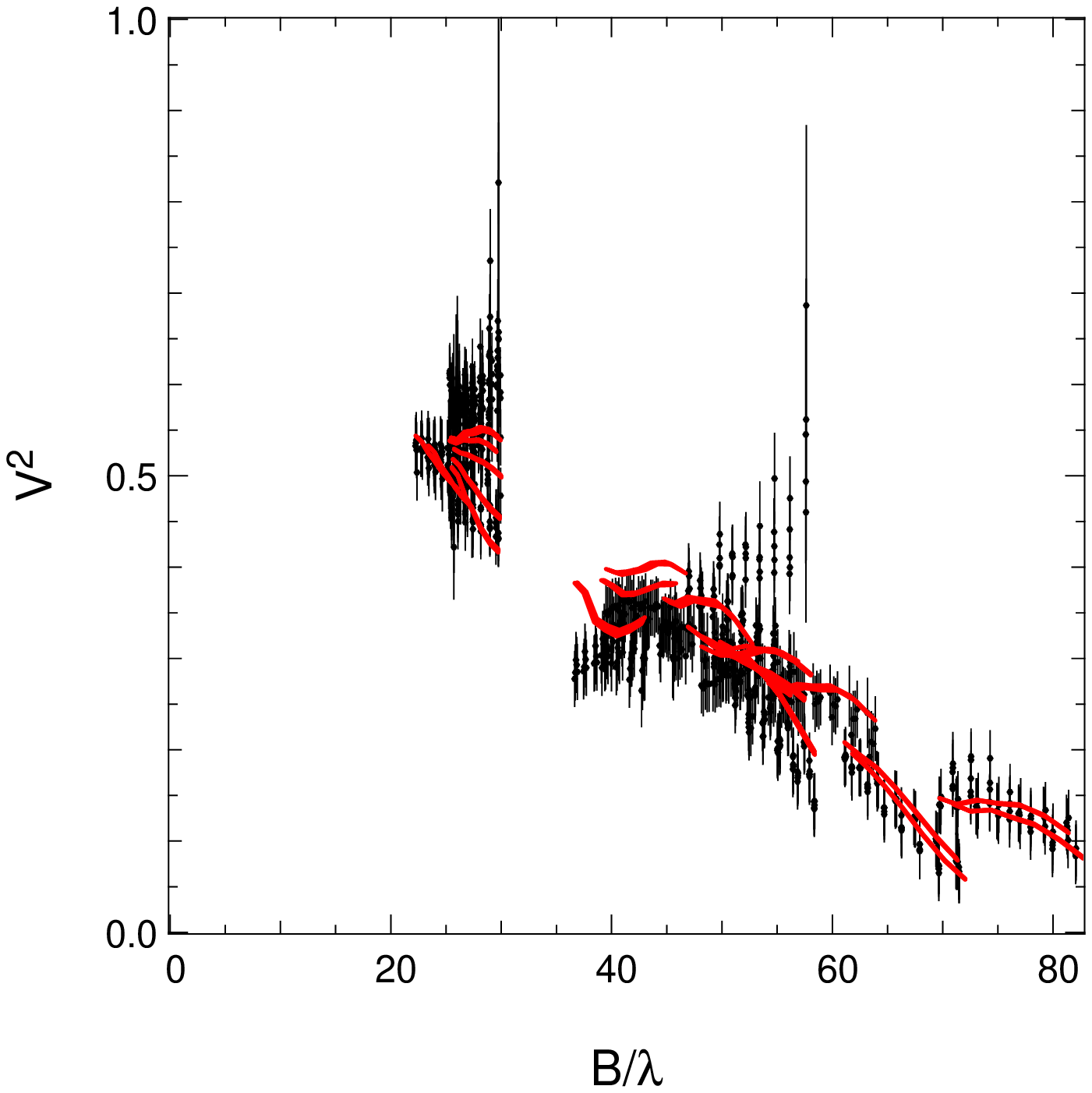}
\caption[]{Visibility curves for the observations of SS Lep. The order of the figures corresponds to the {\it (u,v)}-plan coverages in Fig.\ \ref{fig:UVSSLep}. In black are the data with error bars. The visibility curves from our model with the best parameters are in red. \label{fig:vis2SSLep}}
\end{figure*}

\begin{figure*}[h!]
\centering
\includegraphics[width=0.22\textwidth]{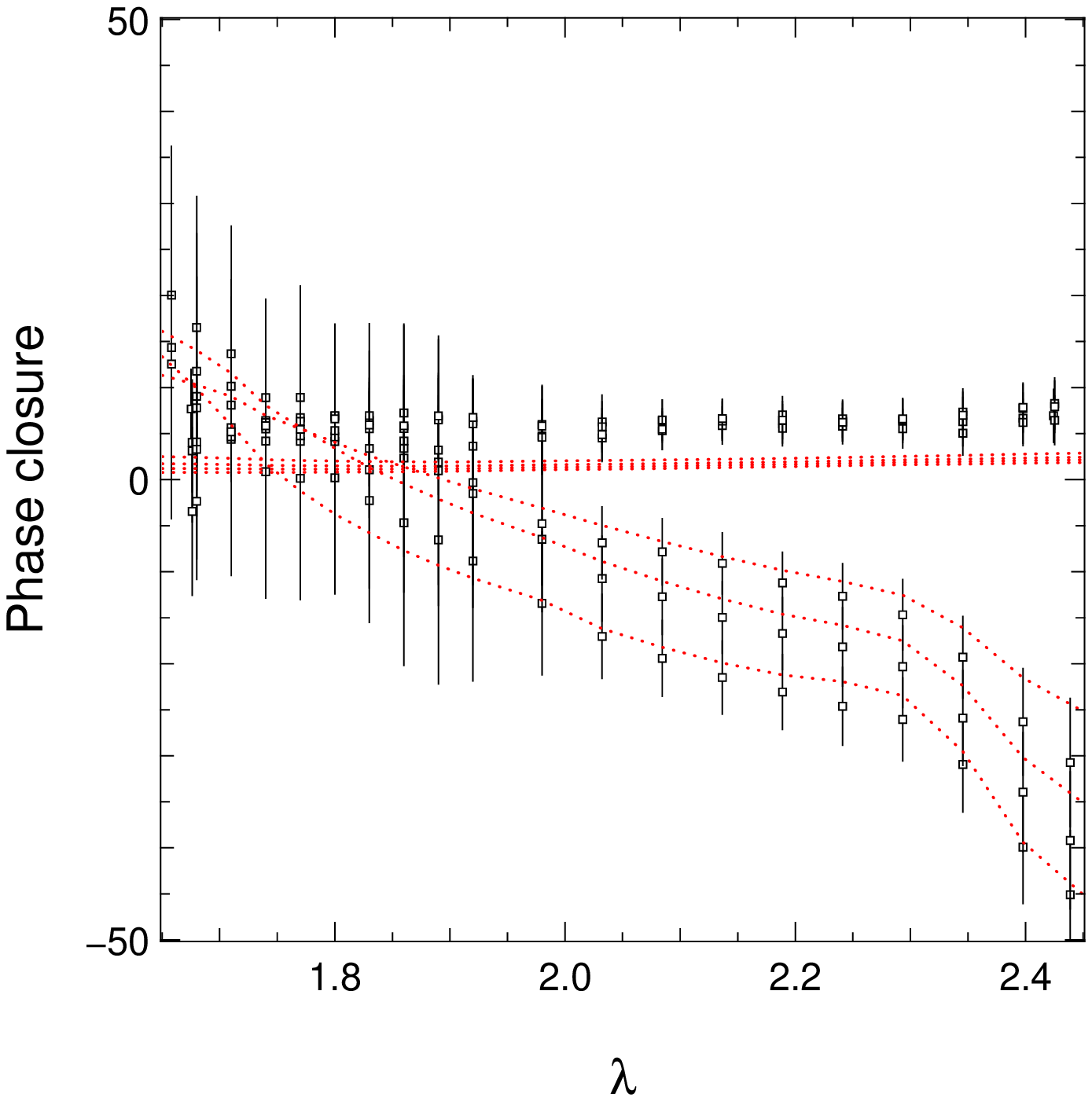}
\includegraphics[width=0.22\textwidth]{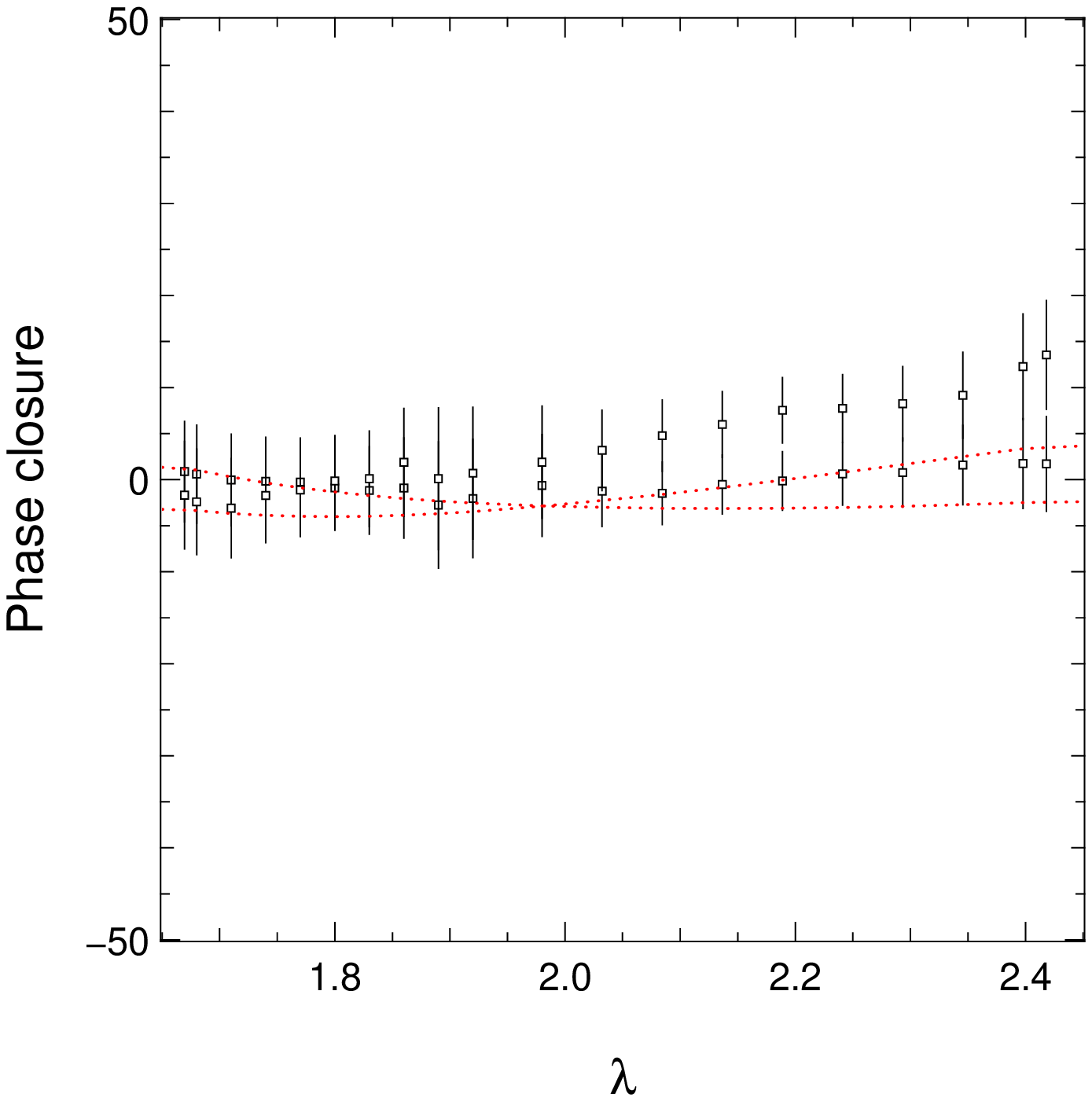}
\includegraphics[width=0.22\textwidth]{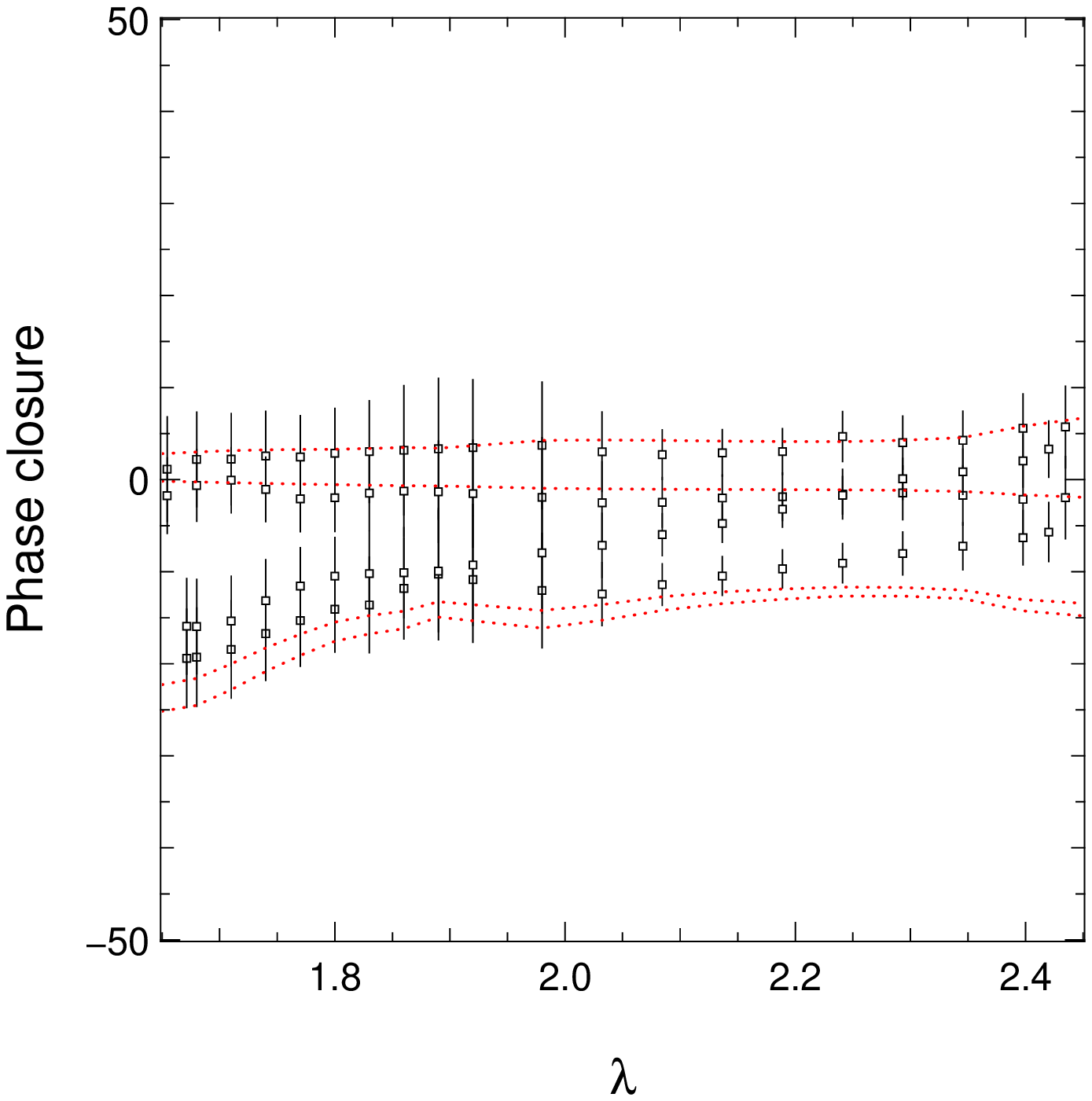}
\includegraphics[width=0.22\textwidth]{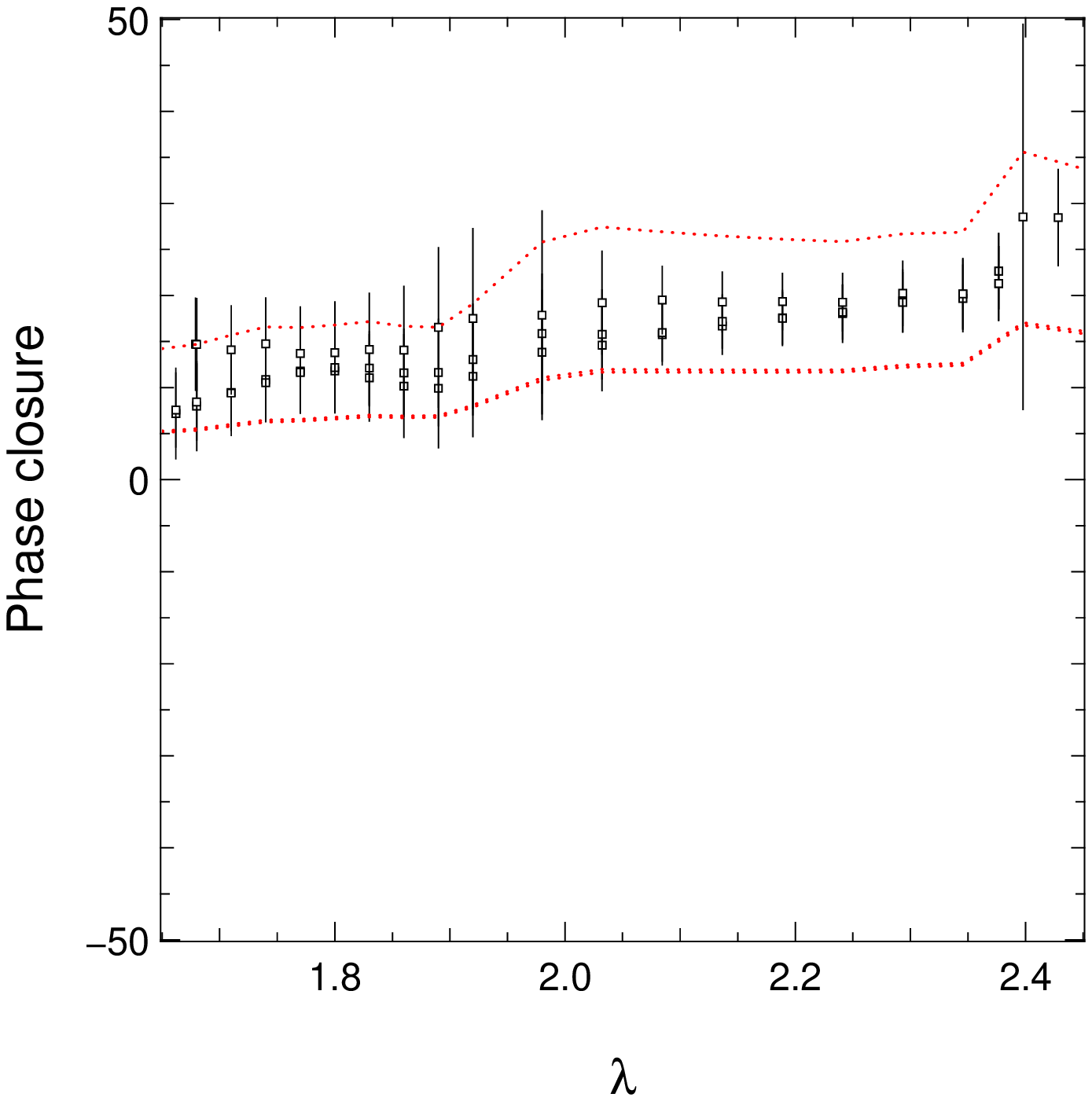}\\
\includegraphics[width=0.22\textwidth]{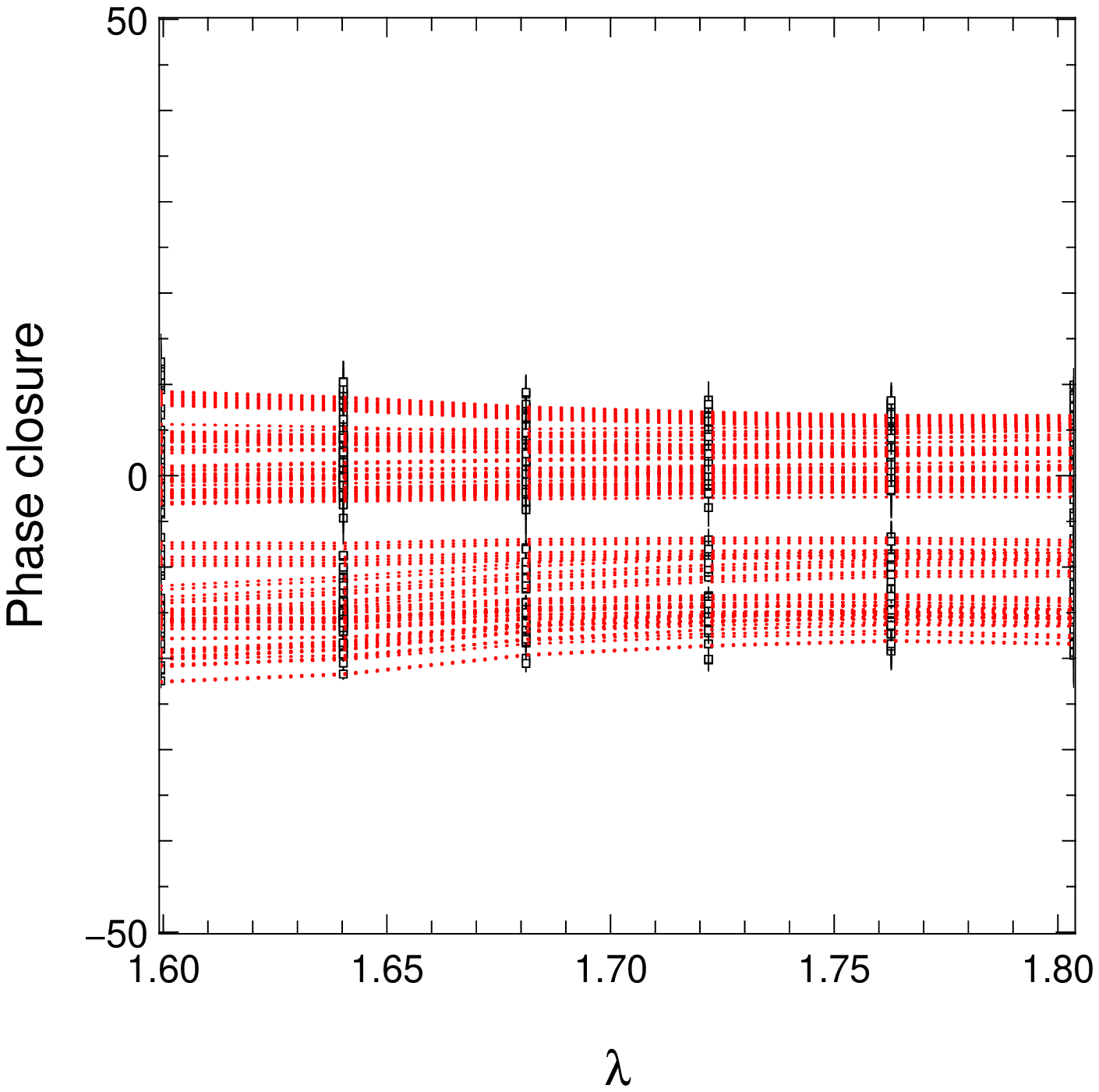}
\includegraphics[width=0.22\textwidth]{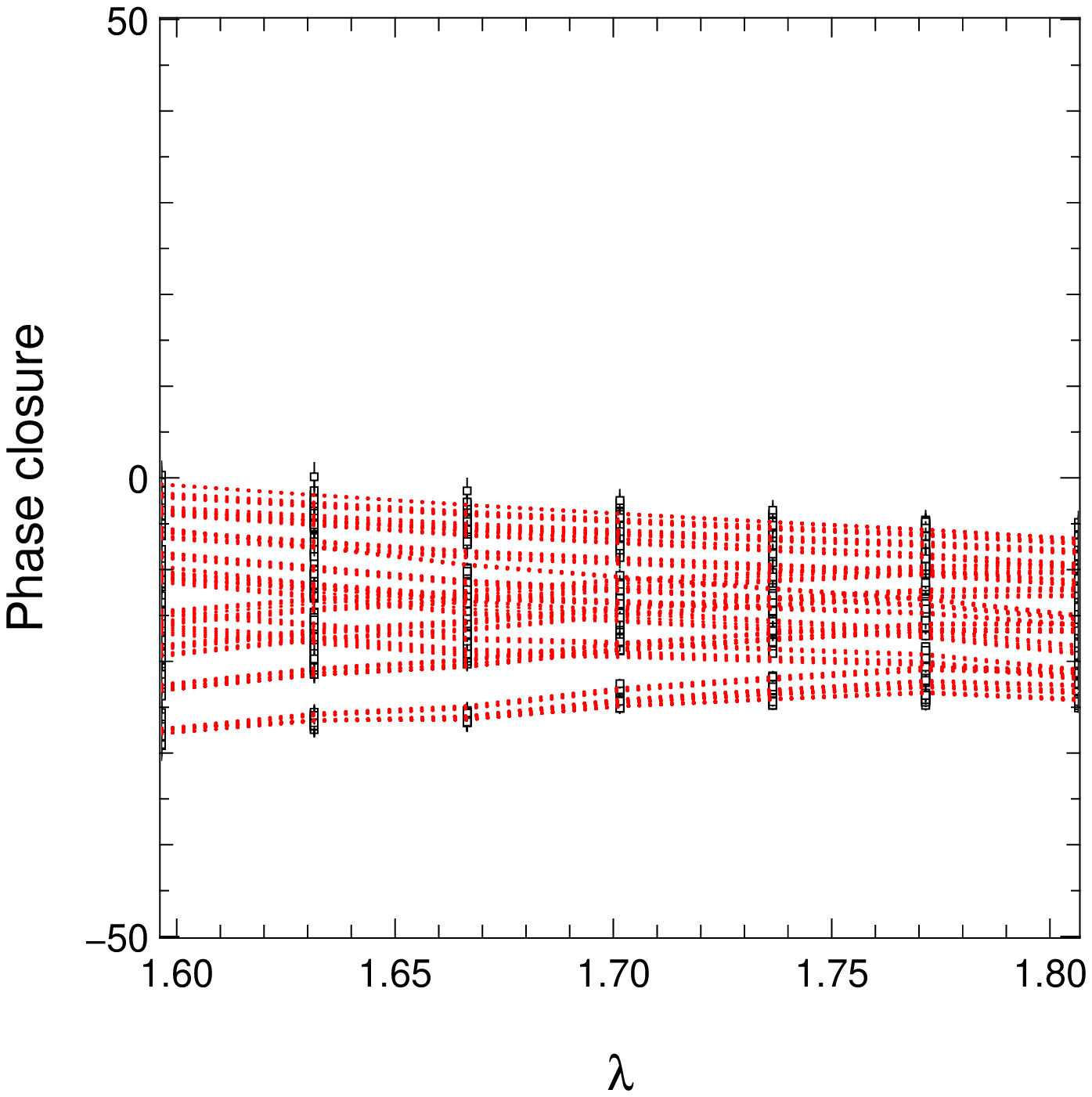}
\includegraphics[width=0.22\textwidth]{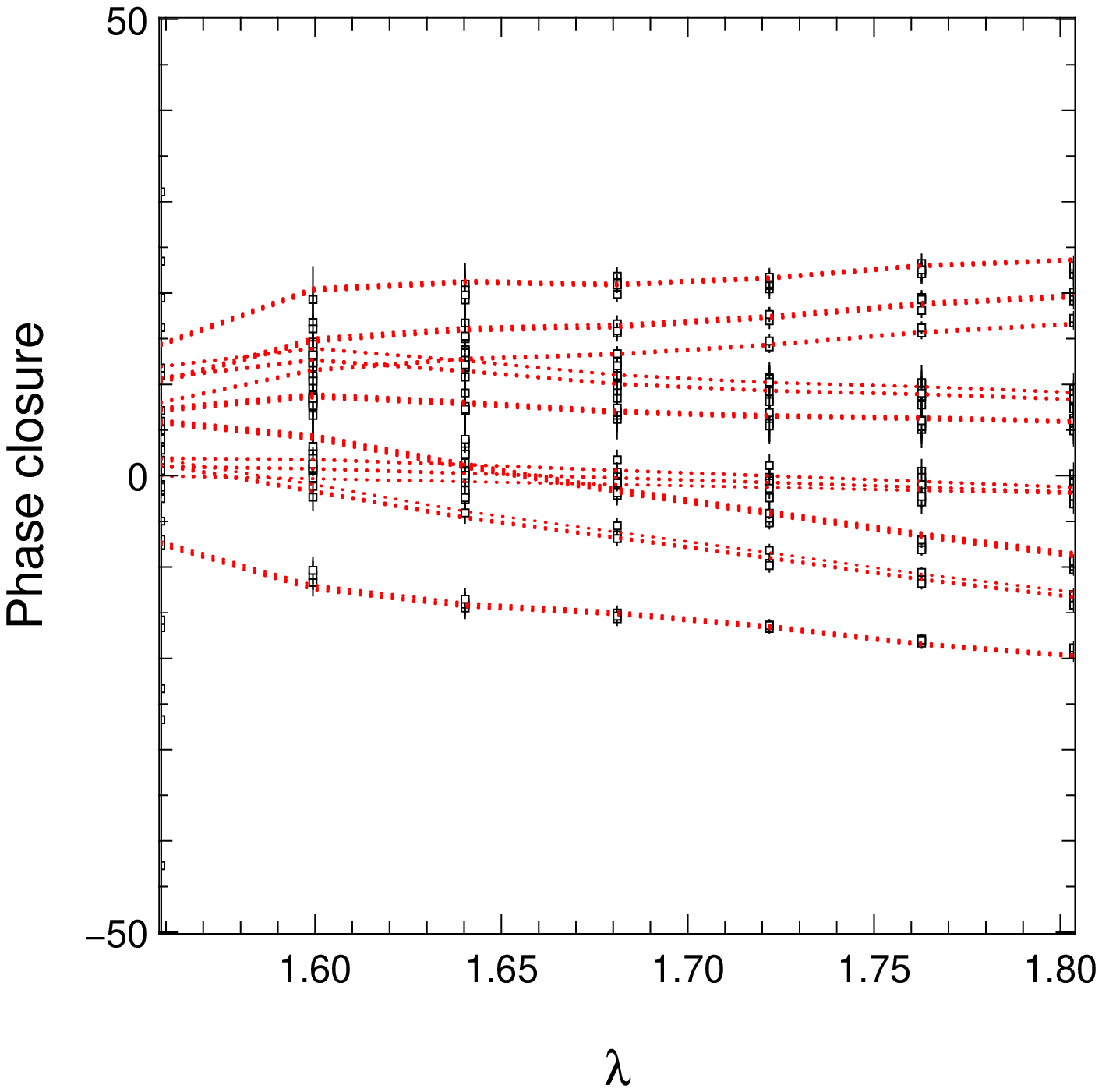}
\includegraphics[width=0.22\textwidth]{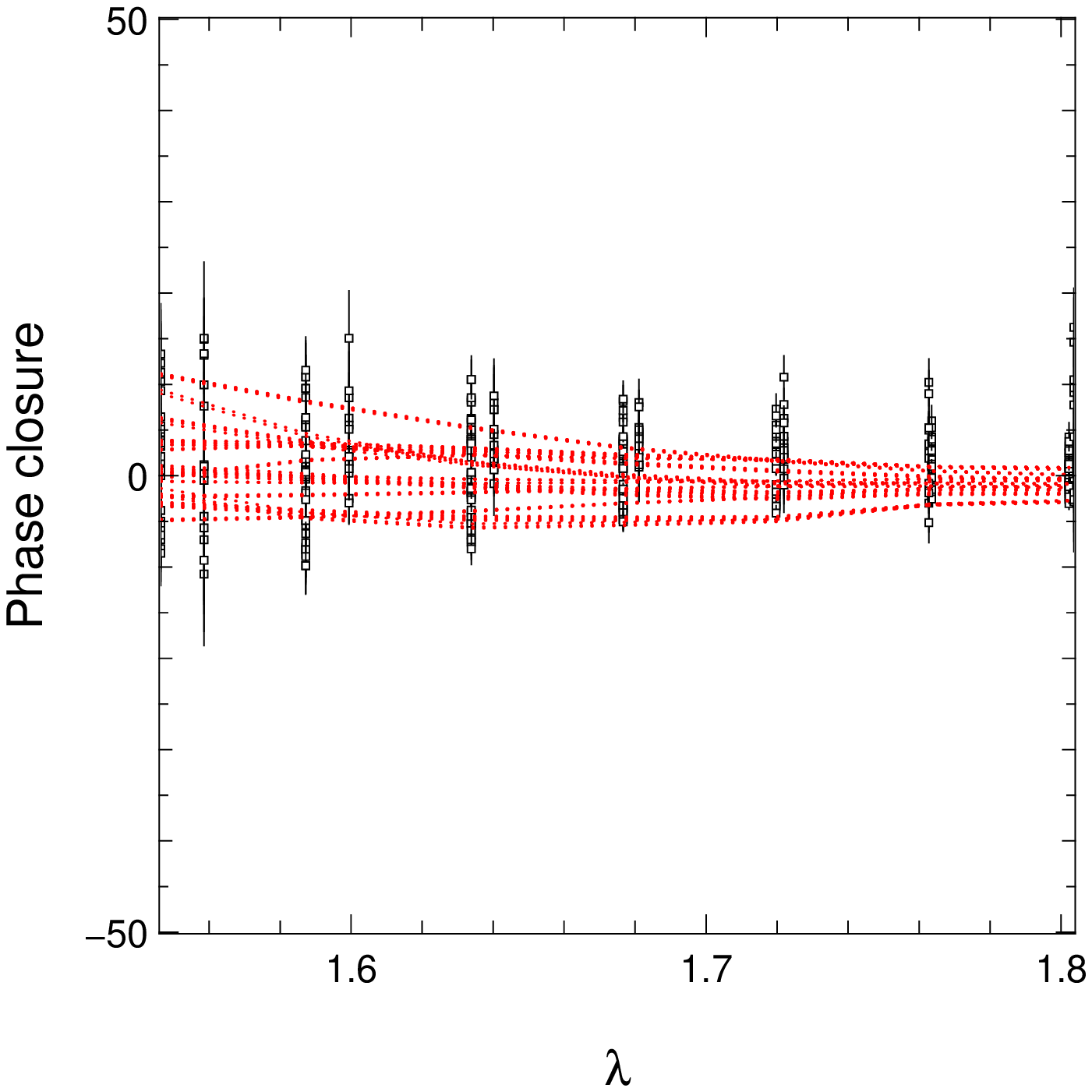}
\caption[]{Closure phase curves for the observations for the observations of SS Lep. The order of the figures corresponds to the {\it (u,v)}-plan coverages in Fig. \ref{fig:UVSSLep} In black are the data with error bars. The visibility curves from our model with the best parameters are in red. \label{fig:CPSSLep}}
\end{figure*}

\begin{figure}[h!]
	\centering
	\includegraphics[width=.4\textwidth]{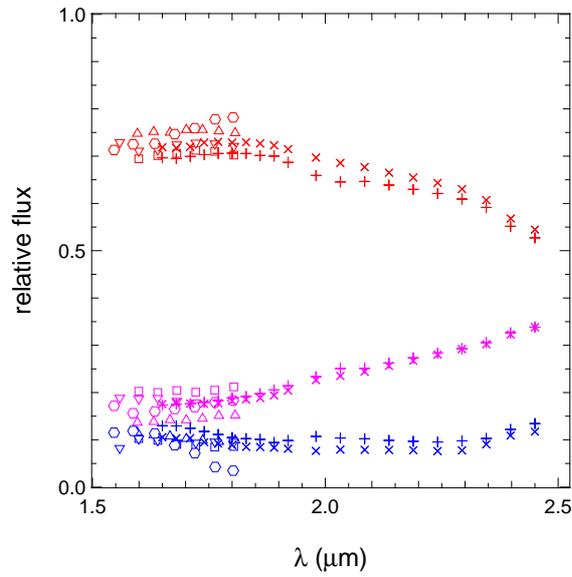}
	\caption{Relative flux contribution of the M giant (red), the A star (blue), and the envelope (magenta). The different symbols represent the different observations A1 ($+$), A3 ($\times$), P1 ($\Box$), P2 ($\bigtriangleup$), P3 ($\bigtriangledown$), P4 ($\bigcirc$). \label{fig:fit_result}}
\end{figure}

The interferometric data are plotted in Figs. \ref{fig:UVSSLep}, \ref{fig:vis2SSLep}, and \ref{fig:CPSSLep} (visibilities and closure phases). The reduced data are available on Vizier/CDS. We also plotted on these figures the visibilities and closure phases obtained from our binary model with the best parameters. Table~\ref{tab:fit_result} and Fig.~\ref{fig:fit_result} present the best parameters with the data. We indicate for each observation which parameters are free in the fitting procedure. The relative fluxes are always free parameters and are considered to be chromatic. Uncertainties on the fitted parameters were computed with Monte-Carlo simulations from the errors on the visibilities and closure phases.

%
%
%
%
%
%

\begin{table}[h!]
\caption{Most reliable parameters obtained with the 8 data sets. } \label{tab:fit_result}
\begin{centering}
	\begin{tabular}{llllllll}
\hline \hline
	& $\Phi_M$ & $\Phi_D$ & $a$ & $\theta$ & $\chi^2_r$\\
	& [mas] & [mas] & [mas] & [deg] & \\
\\
\multicolumn{6}{l}{{\bf Period A1}: $mjd\sim$ 54781 and 54783  (AMBER data)} \\
\hline
Free\,?          & Yes & Yes & Yes & Yes &\\
Value &  $2.24$ & $12.35$ & $4.2$ & $135$ & 0.71\\
Error (1-$\sigma$)& $2.7 \times 10^{-2}$ & $2.3 \times 10^{-1}$ & $3.5 \times 10^{-2}$ & $0.35$ &\\
\\
\multicolumn{6}{l}{{\bf Period A2}: $mjd\sim$ 54826 (AMBER data)} \\
\hline
Free\,?           & Yes & No  & Yes & Yes &\\
Value & $2.05$ & $12.2$ & $3.8$ & $69$ & 0.20\\
Error (1-$\sigma$) & $2.9 \times 10^{-2}$ & - & $4.9 \times 10^{-2}$ & $0.53$ &\\
\\
\multicolumn{6}{l}{{\bf Period A3}: $mjd\sim$ 54883 and 54890  (AMBER data)} \\
\hline
Free\,?          & Yes & Yes & Yes & Yes & \\
Value & $2.14$ & $12.2$ & $4.5$ & $347$ & 1.36\\
Error (1-$\sigma$) & $4.3 \times 10^{-2}$ & $2.3\times 10^{-1}$ & $3.8 \times 10^{-2}$ & $0.37$ &\\
\\
\multicolumn{6}{l}{{\bf Period A4}: $mjd\sim$ 54928  (AMBER data)} \\
\hline
Free\,?        & Yes & No & Yes & Yes & \\
Value & $1.98$ & $12.2$ & $4.2$ & $291$ & 1.62 \\
Error (1-$\sigma$) & $3.9 \times 10^{-2}$ & - & $4.0 \times 10^{-2}$ & $0.47$ &\\
\\
\multicolumn{6}{l}{{\bf Period P1}: $mjd\sim$ 55500 (PIONIER data)} \\
\hline
Free\,?           & Yes & No & Yes & Yes &\\
Value & $2.17$ & $12.2$ & $3.9$ & $32$ &  1.58\\
Error (1-$\sigma$)& $1.4 \times 10^{-2}$ & - & $1.5 \times 10^{-2}$ & $0.06$ &\\
\\
\multicolumn{6}{l}{{\bf Period P2}: $mjd\sim$ 55529 (PIONIER data)} \\
\hline
Free\,?           & Yes & No & Yes & Yes &\\
Value  & $2.27$ & $12.2$ & $4.5$ & $351$ &  2.46\\
Error (1-$\sigma$)& $1.2 \times 10^{-2}$ & - & $0.2 \times 10^{-2}$ & $0.78$ &\\
\\
\multicolumn{6}{l}{{\bf Period P3}: $mjd\sim$ 55537 (PIONIER data)} \\
\hline
Free\,?           & Yes & No & Yes & Yes &\\
Value & $2.19$ & $12.2$ & $4.5$ & $346$ &  0.96 \\
Error (1-$\sigma$)& $1.6 \times 10^{-2}$ & - & $0.6 \times 10^{-2}$ & $0.20$ &\\
\\
\multicolumn{6}{l}{{\bf Period P4}: $mjd\sim$ 55552 (PIONIER data)} \\
\hline
Free\,?           & Yes & No & Yes & Yes &\\
Value & $2.30$ & $12.2$ & $4.42$ & $330.6$ &  0.96 \\
Error (1-$\sigma$) & $0.9 \times 10^{-2}$ & - & $0.9 \times 10^{-2}$ & $0.12$ &\\
	\end{tabular}
	\tablefoot{The errors are the result of Monte Carlo computations based on the error on the visbilities and closure phases measurements.$\phi_M$ and $\phi_D$ are the M star diameter and the dusty disc/envelope FWHM, respectively. $a$ and $\theta$ are the visual separation and orientation of the binary. The last column gives the final reduced $\chi^2$ of the fit.}
\end{centering}
\end{table}
%
%
%

\end{document}